\documentclass[aps,prl,twocolumn,amsmath,amssymb,footinbib,showpacs,longbibliography,superscriptaddress]{revtex4-1}

\usepackage{graphicx,natbib}% Include figure files
\usepackage{dcolumn}% Align table columns on decimal point
\usepackage[mathscr]{euscript}
\usepackage{mathrsfs}
\usepackage{amsmath}
\usepackage{bbold}
\usepackage{bm}% bold math
\usepackage{bbm}
\usepackage{color}
\usepackage{braket}
\usepackage{upgreek}
\usepackage[english]{babel}
\usepackage{latexsym}
\usepackage{subfigure}
\usepackage{epsfig}
\usepackage{hyperref}
\usepackage[T1]{fontenc}
\usepackage[latin9]{inputenc}
\usepackage{amstext}
\usepackage{amssymb}
\usepackage{graphicx}
\usepackage{esint}
\usepackage{babel}

\newcommand{\Jprl}{Phys. Rev. Lett.}

\newcommand{\Jpra}{Phys. Rev. A}
\newcommand{\Jprb}{Phys. Rev. B}

\newcommand{\Jpre}{Phys. Rev. E}
\newcommand{\Jrmp}{Rev. Mod. Phys.}

\newcommand{\lettersection}[1]{\paragraph*{#1.--}}

\begin{document}
\title{A momentum dependent optical lattice induced by artificial gauge potential}

\author{Zekai Chen}
\email[]{zchen57@ur.rochester.edu}
\affiliation{Department of Physics and Astronomy, University of Rochester, Rochester, New York 14627, USA}
\affiliation{Center for Coherence and Quantum Optics, University of Rochester, Rochester, New York 14627, USA}
\author{Hepeng Yao}
\affiliation{
Department of Quantum Matter Physics, University of Geneva, 24 Quai Ernest-Ansermet, CH-1211 Geneva, Switzerland}
\author{Elisha Haber}
\author{Nicholas P. Bigelow}
\email[]{nicholas.bigelow@rochester.edu}
\affiliation{Department of Physics and Astronomy, University of Rochester, Rochester, New York 14627, USA}
\affiliation{Center for Coherence and Quantum Optics, University of Rochester, Rochester, New York 14627, USA}

\date{\today}

\begin{abstract}
	We propose an experimentally feasible method to generate a one-dimensional optical lattice potential in an ultracold Bose gas system that depends on the transverse momentum of the atoms. The optical lattice is induced by the artificial gauge potential generated by a periodically driven multi-laser Raman process, which depends on the transverse momentum of the atoms. We study the many-body Bose-Hubbard model in an effective 1D case and show that the superfluid--Mott-insulator transition can be controlled via tuning the transverse momentum of the atomic gas. We examined our prediction via a strong-coupling expansion to an effective 1D Bose-Hubbard model and a quantum Monte Carlo calculation, and discuss possible applications of our system.
\end{abstract}

\maketitle

%%%%%%%%%%%%%%%%%%%%%%%%%%%%%%%%%
\lettersection{Introduction}

In the past two decades, artificial gauge potentials have become a powerful tool for Hamiltonian engineering in cold atom systems. They enable physicists to use the cold atom ensemble for quantum simulation of complicated condensed matter systems, and lead to interesting physics such as creating novel topological defects \cite{sugawa2018second}, the spin Hall effect (SHE) \cite{zhu2006spin,liu2007optically,leblanc2012observation,beeler2013spin} and spin-orbit coupling (SOC) \cite{dalibard2011colloquium,goldman2014light,zhai2015degenerate,zhang2018spin}. The SHE in a cold atom ensemble---a phenomenon that arises from a spin-dependent Lorentz-like force acting on moving particles from a transverse direction---has drawn significant attention, as it is closely related to quantum Hall physics and spintronics. To date, much of the work studying the SHE in cold atom systems is focused on weakly-interacting single-particle physics such as single-particle spin-dependent trajectories and the SHE induced spin current. On the other hand, SOC---a type of interaction that couples the spin of a particle with its external degree of freedom---is also of extensive interest. It leads to the novel energy dispersion \cite{huang2016experimental,hamner2015spin} and the topological order \cite{lin2011spin,liu2013manipulating,wu2016realization,wang2021realization} in both continuous gas and optical lattice systems. 

Meanwhile, optical lattices with ultracold atoms are well developed platforms for studying quantum phenomena, especially many-body effects of condensed matter systems \cite{bloch2008many}. A familiar aspect of this many-body physics is the superfluid--Mott-insulator (SF-MI) transition, which has been well studied by many groups both theoretically \cite{jaksch1998cold,boeris2016} and experimentally \cite{greiner2002quantum,haller2010,soltan2011multi,boeris2016}. For optical lattice systems, a natural question arises: what many-body effects in an optical lattice system involve an artificial gauge potential? There are a few studies about the many-body effects in a lattice system with a gauge potential \cite{cai2012magnetic,cole2012bose}, but generally the many-body effect caused by either SOC or the SHE in an optical lattice system demand more exploration. In particular, a system where the lattice potential is determined by the motional state of the atoms is of special interest.

In this letter, we propose to realize a novel transverse-momentum-dependent optical lattice (TMDOL) where the longitudinal lattice potential depends on the motional state of the atoms in the transverse dimension. The momentum dependence derives from the same mechanism as the SHE and SOC, which makes it possible to tune the many-body phase of the system via the transverse momentum of the ultracold gas. We create such an artificial gauge potential with translational symmetry in an ultracold pseudospin-1/2 gaseous system via a periodically driven Raman process. To simplify the problem, we project the Hilbert space to a single spin component and explore the many-body physics of the TMDOL with the help of a strong transverse confinement. We construct the effective 1D Bose-Hubbard model, and investigate the SF-MI phase diagram using a strong coupling expansion. We then examine our Bose-Hubbard model result with a quantum Monte Carlo (QMC) calculation. Both calculations show that the SF-MI phase transition can be induced by changing the average transverse momentum in such a TMDOL, which arises as a novel many-body effect caused by the artificial gauge potential. In addition, we propose a feasible experimental implementation of the TMDOL and discuss the possible applications.

%%%%%%%%%%%%%%%%%%%%%%%%%%%%%%%%%
\lettersection{Theoretical formalism}

We first describe the theoretical construction of the optical lattice for a pseudospin-1/2 system with a periodically driven Hamiltonian. We start by considering the Hamiltonian
\begin{equation}\label{construction Hamiltonian}
H=\frac{\vec{p}^2}{2M}+V_{ext}(\vec{r})+H_{d},
\end{equation}
where $M$ is the mass of the particle, $V_{ext}(\vec{r})$ is the spin-independent external trapping potential and $H_d$ is a periodically driven Hamiltonian $H_{d}(\vec{r},t)=\frac{\hbar}{2}\Omega_0\vec{n}\cdot\vec{\sigma}\cos{\omega t}$, with $\vec{n}=(\cos{k_Lx}\cos{k_Lz},\cos{k_Lx}\sin{k_Lz},0)^T$ the direction vector, $k_L$ the wave vector, $\Omega_0$ the amplitude of the laser field, $\vec{\sigma}=(\hat{\sigma}_1,\hat{\sigma}_2,\hat{\sigma}_3)^T$ the vector of Pauli matrices, and $\omega$  the frequency of the periodic driving. According to Floquet theory, we can apply a gauge transformation to the Hamiltonian described by the micromotion operator $U=e^{-i\sin{\omega t}\vec{\Omega}(r)\cdot\vec{\sigma}/2\omega}$ \cite{ravckauskas2019non,novivcenko2019non,chen20202}, where $\vec{\Omega}=\Omega_0\vec{n}$ and the Hamiltonian in the Floquet basis can be written as
\begin{equation}\label{Floquet Hamiltonian}
H_{F}(r,t)=\frac{\left[\vec{p}-\vec{A}(r,t)\right]^2}{2M}+V_{ext}(\vec{r}),
\end{equation}
where $\vec{A}(r,t)=i\hbar U^{\dagger}\nabla U$ is the non-Abelian gauge potential, with the specific form of the component given in the supplemental material \cite{supplemental}. We can ignore all terms higher than the zeroth-order in the Fourier series if the adiabatic condition, $|\bra{\vec{k}} H_{F}^{(n)} \ket{\vec{k}^{'}}|\ll\hbar\omega (n\ne0)$, is satisfied \cite{ravckauskas2019non,novivcenko2019non,supplemental}, where $H_{F}^{(n)}=\frac{1}{\mathcal{T}}\int_{0}^{\mathcal{T}}H_{F}(t)e^{-in\omega t}dt$ is the $n$-th Fourier component of the Hamiltonian in the Floquet basis, $\mathcal{T}=2\pi/\omega$, and $\ket{\vec{k}}$ is the momentum state with momentum $\hbar\vec{k}$.

The gauge field exerts a spatially periodic `Lorentz' force on atoms along the $x$-axis due to the motion of atoms along the $z$-axis, which makes the lattice potential $p_z$-dependent. The non-zero component of the zeroth Fourier order of the gauge potential takes the form $A_{z}^{(0)}=\hbar k_L[J_0(a)-1]\sigma_3/2$, where $J_0(a)$ is the zeroth-order Bessel function of the first kind and $a=a_0|\cos{k_Lx}|$ with $a_0=\Omega_0/\omega=4$ throughout this paper. Also we get $(\vec{A}^2)^{(0)}=\hbar^2k_L^2\{\frac{1}{16}a_0^2(1-\cos{2k_Lx})+\frac{1}{2}[1-J_0(a)]\}$. Keeping only the zeroth Fourier component of the gauge potential in eqn.(\ref{Floquet Hamiltonian}) and projecting to spin up subspace, we get the effective Hamiltonian in the Floquet basis as
\begin{equation}\label{single spin lattice Hamiltonian}
H_{L}=\frac{\vec{p}^2}{2M}+V(x,p_z)+V_{ext}(\vec{r}),
\end{equation}
where $E_r=\hbar^2k_L^2/2M$ is the recoil energy. $V(x,p_z)=\{(p_z+1/2)\left[1-J_0(a)\right]+a_0^2(1-\cos{2k_Lx})/16\}E_r$ is an optical lattice potential with periodicity $a_L=\pi/k_L$. In this work, we only consider the case with $a_0=4$. Eqn.(\ref{single spin lattice Hamiltonian}) shows the $p_z$-dependence of the lattice potential, which results in $p_z$-dependent ground states \cite{supplemental}. 

%%%%%%%%%%%%%%%%%%%%%%%%%%%%%%%%%
\lettersection{Many-body Hamiltonian}

We desire to study the many-body effects in our system with a non-negligible interaction strength. Therefore, we introduce a deep two-dimensional optical lattice confinement along the $y$ and $z$-axes so that the tunneling along $y$ and $z$ is negligible and the system can be regarded as an array of effective 1D Bose gases. With a tight 2D transverse confinement along the $y$ and $z$-axes, the ultracold gas will be trapped in an array of effective 1D tubes. Throughout this paper, we consider the case where the transverse confinement is strong enough that the tunneling between nearest sites in the $y$ and $z$ is negligible \cite{stoferle2004transition,anisimovas2016semisynthetic} (e.g. a lattice depth $V_{\perp}=30E_r$). Furthermore, we consider a homogeneous lattice along the $x$-axis for simplicity. The transverse trapping potential for each 1D tube can be well approximated by a harmonic trap with the trapping frequency $\omega_{\perp}=2\sqrt{V_{\perp}E_{r,\perp}}/\hbar \gg \mu$, where $\mu$ is the chemical potential, and $V_{\perp}$ and $E_{r,\perp}=\hbar^2k_{\perp}/2M$ are the lattice depth and the recoil energy of the 2D lattice. One can separate the dimensions and the ground state wavefunctions along the $y$ and $z$-axes can be well approximated by the harmonic ground state wave packets $\varphi(y)=(\sqrt{\pi}a_{\perp})^{-\frac{1}{2}}exp\{-\frac{y^2}{2a_{\perp}^2}\}$ and $\varphi(z)=(\sqrt{\pi}a_{\perp})^{-\frac{1}{2}}exp\{-(\frac{z^2}{2a_{\perp}^2}+i\bar{k}_zz)\}$,
where $a_{\perp}=\sqrt{\hbar/M\omega_{\perp}}$ is the characteristic length of the tight harmonic trap and $\bar{k}_z$ corresponds to the average momentum along $z$-axis, $\langle p_z\rangle=\hbar \bar{k}_z$.

Due to the uncertainty principle, the tight transverse confinement gives rise to a non-negligible spreading in momentum space of the atoms along $y$ and $z$-axes. By integrating along the $y$ and $z$-axes, one gets the single-particle effective 1D Hamiltonian \cite{barbiero2014quantum}
\begin{eqnarray}\label{effective 1D Hamiltonian single particle}
H_{1D}(x,\bar{k}_z)&=&\int\int dydz \varphi^{\ast}(y) \varphi^{\ast}(z)H_{L}\varphi(z)\varphi(y) \nonumber \\
&=&\frac{p_x^2}{2M}+\sum_{k}|\tilde{\varphi}(k-\bar{k}_z)|^2V(x,p_z),
\end{eqnarray}
where $\tilde{\varphi}(k-\bar{k}_z)$ is the Fourier transform of $\varphi(z)$, $V_{1D}(x,\bar{k}_z)=\sum_{k_1}|\tilde{\varphi}(k_1-\bar{k}_z)|^2V(x,p_z)$ is the effective 1D potential and can be written in the form of
\begin{equation}\label{V1D}
    V_{1D}(x,\bar{k}_z)=\sum_{m=0}^{\infty}V_{m}(\bar{k}_z)\cos{2mk_Lx},
\end{equation} where $V_{m}(\bar{k}_z)$ is the $m$-th order coefficient that depends on $\bar{k}_z$. We find truncating at $m=4$ sufficient with $a_0=4$, as the coefficients $V_m$ with $m \geq 3$ are much smaller than the leading orders. Notice that we ignored the constant energy offset, $\hbar\omega_{\perp}+\hbar^2\bar{k}_z^2/2M$.

Next, we consider the 3D many-body Hamiltonian
$\mathcal{H}_{3D}=H_0+H_{int}$, where $H_0=\int d^3r\hat{\psi}^{\dagger}(r)H_L\hat{\psi}(r)$ and $H_{int}=\frac{g}{2}\int d^3r\int d^3r'\hat{\psi}^{\dagger}(r)\hat{\psi}^{\dagger}(r')\delta(r-r')\hat{\psi}(r')\hat{\psi}(r)$.
The bosonic annihilation operator at $j$th lattice site is defined as $\hat{\psi}(r)=\sum_{j}W_{j}(x)\varphi(y)\varphi(z)\hat{a}_{j}$ and $g=4\pi\hbar^2a_s/M$, where $a_s$ is the $s$-wave scattering length. $W_{j}(x)$ and $\hat{a}_{j}$ is the Wannier function and dimensionless annihilation operator of $j$th site for the effective 1D Hamiltonian, $H_{1D}$. We use the Gaussian approximation of the Wannier functions of each site for the lattice potential, $V_{1D}(x,\bar{k}_z)$. Combining eqn.(\ref{effective 1D Hamiltonian single particle}) with the 3D many-body Hamiltonian, we obtain the many-body effective 1D Hamiltonian. Noticing that the next-nearest-neighbor tunneling is negligible in a certain regime of $\langle p_z\rangle=\hbar\bar{k}_z$, we can apply the tight-binding approximation and obtain the effective 1D Bose-Hubbard Hamiltonian,
\begin{eqnarray}\label{BH Hamiltonian}
H_{BH}=&&-\sum_{j}J_{1D}(\hat{a}_{j+1}^{\dagger}\hat{a}_{j}+\hat{a}_{j+1}\hat{a}_{j}^{\dagger})\nonumber \\&&+\frac{U_{1D}}{2}\sum_{j}\hat{n}_{j}(\hat{n}_{j}-1)-\sum_{j}\tilde{\mu}\hat{n}_{j},
\end{eqnarray}
where $\hat{n}_{j}=\hat{a}_j^{\dagger}\hat{a}_j$ is the number operator of $j$th site and $\tilde{\mu}_j=\mu-\varepsilon_{j}$. The effective tunneling, $J_{1D}$, on-site interaction, $U_{1D}$, and energy offset, $\varepsilon_{j}$, take the forms
\begin{eqnarray}\label{J_U_E}
&&J_{1D}(\bar{k}_z)=-\int dx W_{j}^{\ast}(x)H_{1D}(x,\bar{k}_z)W_{j+1}(x), \nonumber \\
&&U_{1D}(\bar{k}_z)=g_{1D}\int dx |W_{j}(x)|^4, \\
&&\varepsilon_{j}(\bar{k}_z)=-\int dx W_{j}^{\ast}(x)H_{1D}(x,\bar{k}_z)W_{j}(x), \nonumber
\end{eqnarray}
where $g_{1D}=2\hbar\omega_{\perp}a_s/(1-Aa_s/a_{\perp})$ with $A=1.036$ \cite{bloch2008many}.
We can drop the site index $j$ so that $\tilde{\mu}_j$ becomes $\tilde{\mu}$, since the lattice is assumed isotropic. 
\begin{figure}
	\begin{center}
		\includegraphics[scale=0.45]{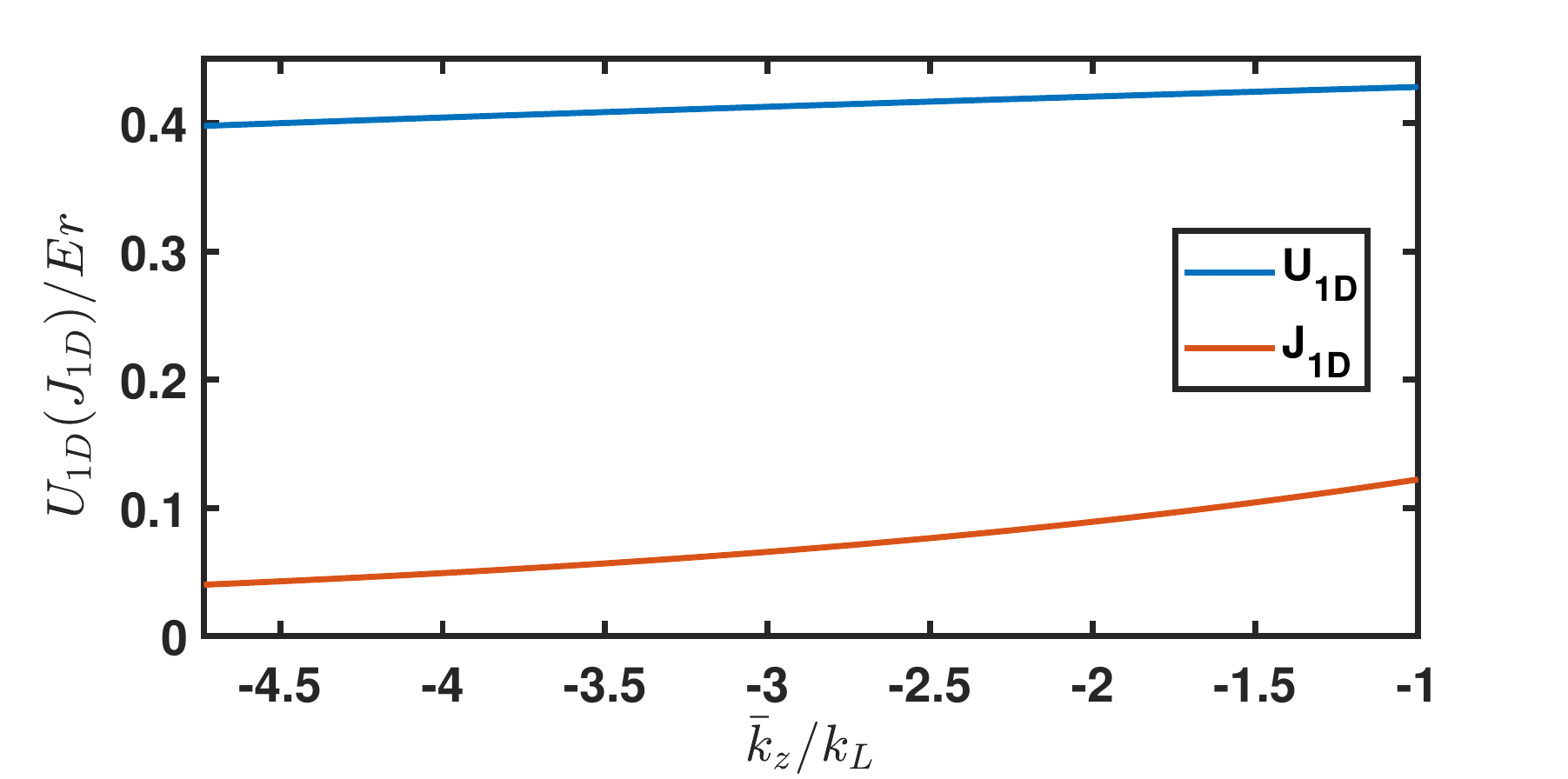}
	\end{center}
	\caption{Effective 1D on-site interaction, $U_{1D}$ (blue line), and effective nearest-neighbor tunneling, $J_{1D}$ (red line), with respect to the average transverse momentum, $\hbar \bar{k}_z$.}\label{parameter changes with pz}
\end{figure}
Eqn.(\ref{J_U_E}) shows that both $J_{1D}$ and $U_{1D}$ depend on the average transverse momentum, $\hbar\bar{k}_z$. In Fig.\ref{parameter changes with pz}, we show the plot of $J_{1D}$ and $U_{1D}$ as functions of average transverse momentum.

%%%%%%%%%%%%%%%%%%%%%%%%%%%%%%%%%
\lettersection{Phase diagram}

\begin{figure}
	\begin{center}
		\includegraphics[scale=0.45]{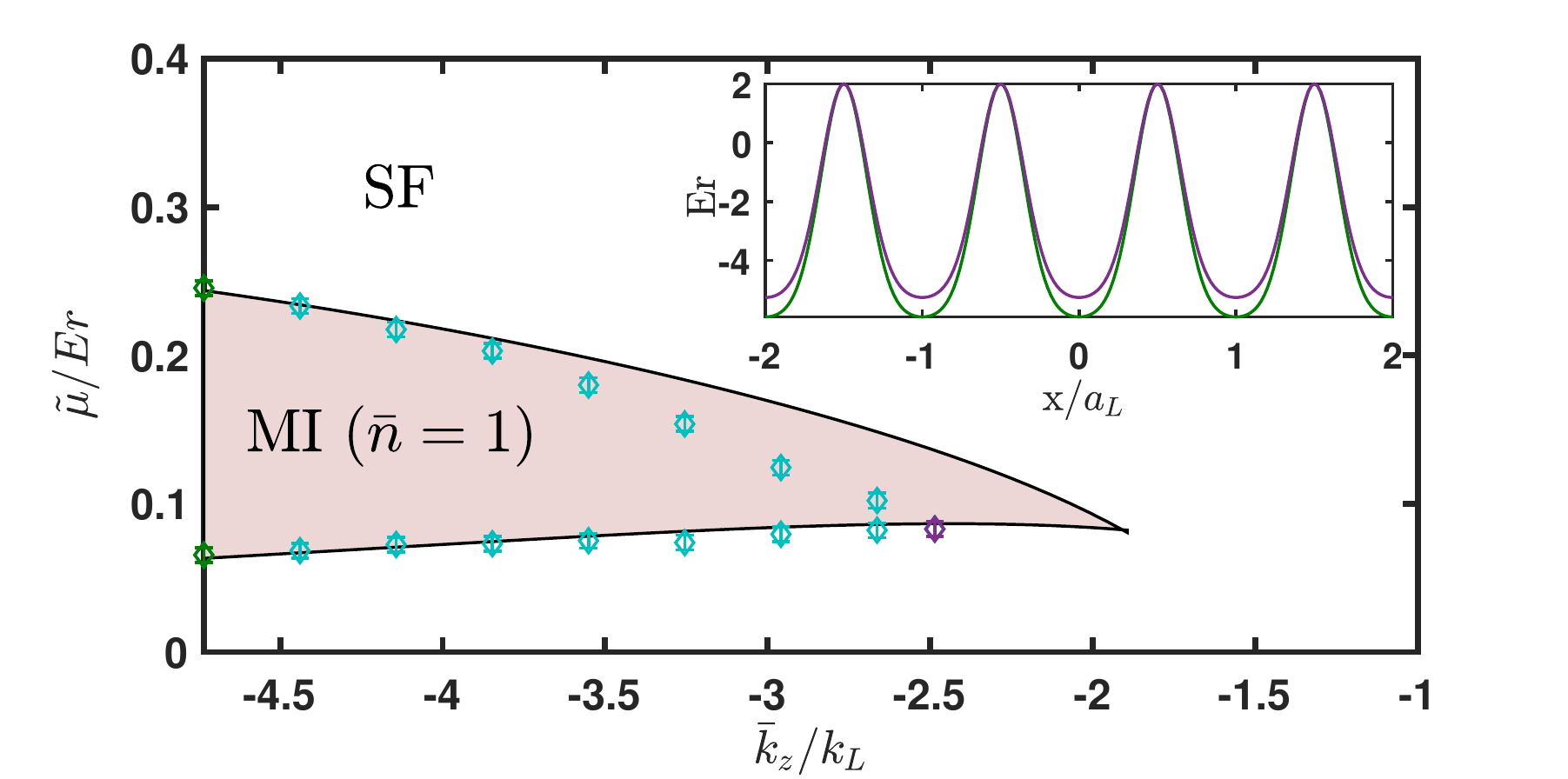}
	\end{center}
	\caption{Phase diagram of the effective 1D TMDOL. The pink shaded area with a black boundary is the $\bar{n}=1$ Mott lobe calculated from the strong coupling expansion of the effective 1D Bose-Hubbard model. Diamonds with errorbars are QMC results with system size $L=50a_L$ and temperature $k_BT=0.004 E_r$. Note that all errorbars are $\pm0.005E_r$. The inset plot shows the effective 1D lattice potential for $\bar{k}_z=-4.73k_L$ (green line) and $\bar{k}_z=-2.49k_L$ (violet line), which correspond to the potentials for the green diamond and the violet diamond data points, respectively. Blue diamonds with errorbars are QMC results with the other momentum.}\label{phase diagram}
\end{figure}

In the effective 1D regime, the on-site interaction strength is much larger than the nearest-neighbor tunneling so that we can use the strong coupling expansion to calculate the SF-MI phase diagram of the Bose-Hubbard model \cite{kuhner1998phases,ejima2012characterization,wang2018high} and get a $\langle p_z\rangle$-dependent phase diagram. The solid line in Fig.\ref{phase diagram} shows the $\bar{n}=1$ Mott lobe of the phase diagram. The phase diagram displays the tunability of the SF-MI phase transition upon the transverse average momentum $\langle p_z\rangle$ in TMDOL. The detailed parameters we use are chosen for $^{87}$Rb atoms. Specifically, $k_L=2\pi/\lambda_R$, $\lambda_R\approx791\textrm{nm}=2a_L$, $k_{\perp}=2\pi/\lambda_{\perp}$, and $\lambda_{\perp}=852\textrm{nm}$ is the wavelength for the 2D optical lattice along transverse directions.

Next, we use path-integral Monte Carlo calculations in continuous space in the grand-canonical ensemble \cite{ceperley1995} to obtain the QMC phase diagram. We consider the effective 1D many-body Hamiltonian in the continuous space,
%-------------------------------%
\begin{equation}\label{eq:Hamiltonian}
H_{QMC} =\sum_{1 \leq j \leq N} \Big[-\frac{\hbar^2}{2m}\frac{\partial^2}{\partial x_j^2}+V_{1D}(x_j,\bar{k}_z)\Big]+g_{1D}\sum_{j<\ell}\delta(x_j-x_\ell),
\end{equation}
%-------------------------------%
where $V_{1D}(x,\bar{k}_z)$ is the effective potential with the form of eqn.(\ref{V1D}).
For a given temperature, $T$, chemical potential, $\mu$, and external potential, $V_{1D}(x,\bar{k}_z)$, we can generate the imaginary-time Feynman diagram configurations with a certain probability distribution. Then, the particle density, $\bar{n}=N/L$, is obtained from the statistics of the closed worldlines, where $N$ denotes the atom number in the ensemble and $L$ the system size. Thanks to the worm algorithm implementations~\cite{boninsegni2006a,boninsegni2006b}, we can further compute the superfluid density, $\rho_{s}$, efficiently. These two quantities help us to distinguish the MI from the SF phase. We use the same QMC algorithm as in Refs.~\cite{boeris2016,yao2018,yao2020,gautier2021}, where details of the implementations are discussed. In the actual QMC calculations, we take a large enough size, $L=50a_L$, and small enough temperature, $k_B T=0.004 E_r$, to make sure our system is in thermodynamic and zero-temperature limits. We confirm that there is no further finite-size and finite-temperature effects at larger sizes and smaller temperatures.

Our final QMC results are shown in Fig.\ref{phase diagram} as the diamonds with errorbars, which indicates the SF-MI transition boundaries for corresponding parameters. We generally find good agreements between the strong coupling expansion and the QMC calculation for the $\bar{n}=1$ Mott lobe. Despite the small difference for small $|k_z|$, both methods confirms a significant Mott lobe for $\bar{n}=1$, which verifies that there exists a TMD SF-MI transition. As $\bar{k}_z$ gets closer to zero, the strong coupling expansion and the QMC results start to deviate from each other. As shown by the inset plot of Fig.\ref{phase diagram}, the lattice potential becomes shallower when $|\bar{k}_z|$ decreases, which leads to the increase of $J_{1D}/U_{1D}$, and the next-nearest neighbor tunneling becomes non-negligible. Therefore, the Bose-Hubbard model becomes less accurate and the deviation between two methods arises. Moreover, the $\bar{n}=2$ Mott lobe is shown in the supplemental material \cite{supplemental}, and we believe that the bigger deviation between two methods is due to the density-dependent ground state \cite{luhmann2012multi,dutta2011bose,bissbort2012effective}. Despite the slight differences, generally the TMD SF-MI transition is still confirmed by both approaches, which displays the many-body effect caused by the artificial gauge potential in TMDOL.

\begin{figure}
	\begin{center}
		\includegraphics[scale=0.5]{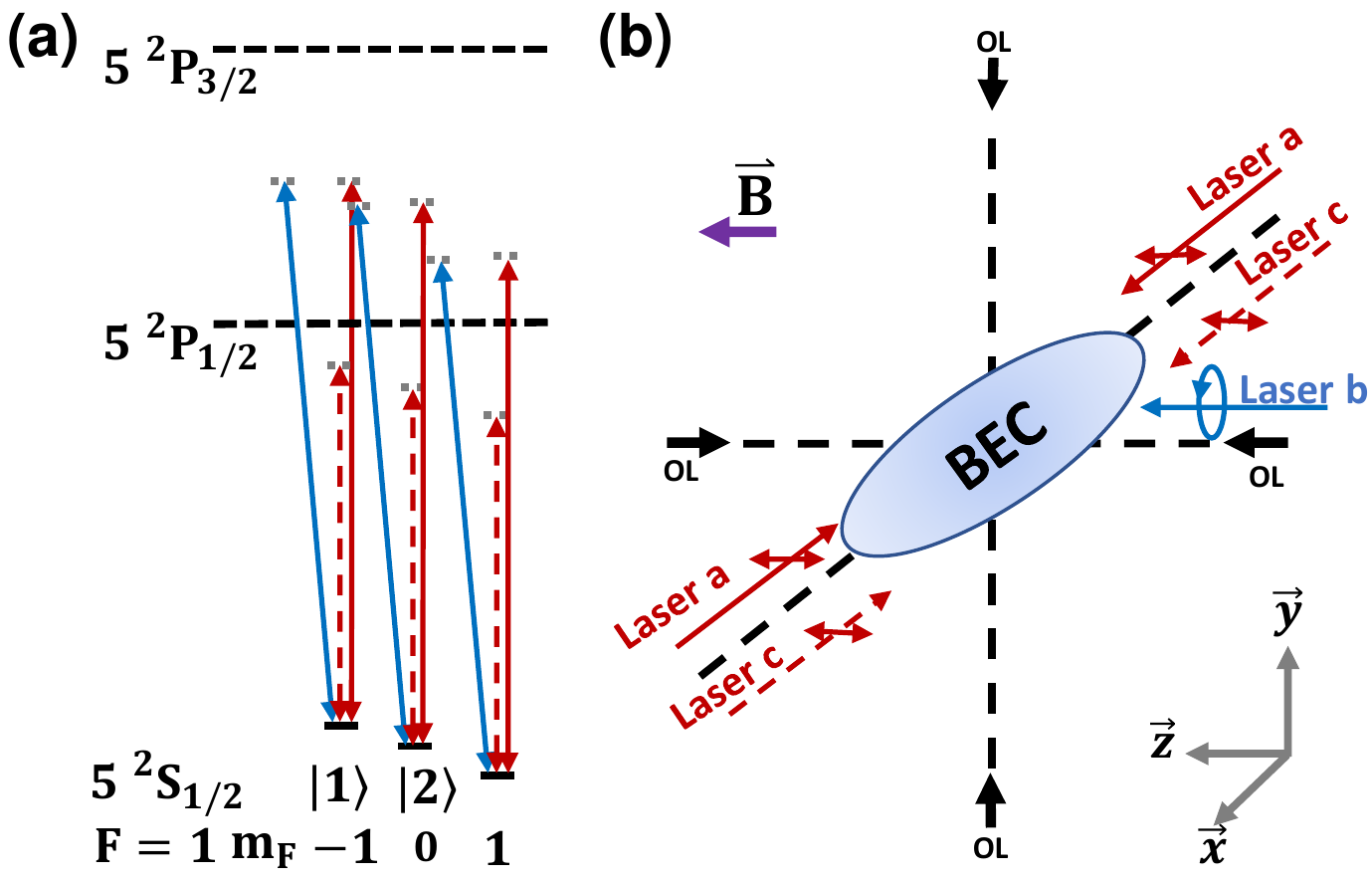}
	\end{center}
	\caption{(a) Level diagram of the Raman process. Red solid, blue solid and red dashed arrows indicate lasers $a$, $b$, and $c$, respectively. (b) Laser configuration of the Raman process. The laser colors are the same as (a), with their polarizations indicated by the small arrows. The bias magnetic field is in the $+z$ direction, and the four black arrows represent the two-dimensional trapping lattice beams in $y-z$ plane.}\label{level diagram}
\end{figure}

%%%%%%%%%%%%%%%%%%%%%%%%%%%%%%%%%
\lettersection{Experimental implementation}

Experimentally, the periodically driven Hamiltonian in eqn.(\ref{construction Hamiltonian}) can be constructed by coupling the atoms to laser fields. Specifically, we consider the dipole interaction of $^{87}$Rb atom with continuous weak (CW) laser fields \cite{wright2008raman}. The pseudospin-1/2 system is formed by two Zeeman states in the $5^{2}S_{1/2}, F=1$ ground state manifold, namely $\ket{\uparrow}=\ket{F=1,m_F=-1}$, and $\ket{\downarrow}=\ket{F=1,m_F=0}$. In the presence of a bias magnetic field, $\vec{B}$, which defines the $z$-axis of the system, the degeneracy between the two states is lifted. Both states are coupled via a Raman process by applying CW Raman lasers $a$, $b$ and $c$, where $a$ and $c$ are standing waves along the $x$-axis with $\pi$-polarization and $b$ propagates in the $+z$-direction with $\sigma^{-}$-polarization. The Raman lasers are detuned far enough from the $D_1$ and $D_2$ transitions to avoid any excitations over the timescale of the experiment. The states we considered in our calculations and the laser configurations are shown in Fig.\ref{level diagram}. Laser coupling to all other states will be negligible.

The sinusoidal driving of the Hamiltonian can be achieved by using a elctro-optical modulator (EOM) to drive the intensity of Raman lasers $a$ and $c$ with $|\cos{\omega t}|$, and at the same time modulating the relative phase, $\eta$, between lasers $a$ and laser $b$ such that $\eta(t)=\pi\left[1-sgn(\cos{\omega t})\right]$, where $sgn(\cdot)$ denotes the signum function. The electric fields of the Raman lasers take the form
\begin{eqnarray}\label{electric field of raman lasers}
&&\vec{E}_a=E_a\hat{\varepsilon}_{0}\cos{k_Lx}|\cos{\omega t}|e^{-i\left[\omega_a t+\eta(t)\right]}  \nonumber \\
&&\vec{E}_b=E_b\hat{\varepsilon}_{-}e^{-i(k_Lz+\omega_b t)}  \\
&&\vec{E}_c=E_c\hat{\varepsilon}_{0}\cos{k_Lx}|\cos{\omega t}|e^{-i\omega_c t}, \nonumber
\end{eqnarray}
where $E_j$ and $\omega_j$ ($j=a,b,c$) is the amplitude and the angular frequency of electric field of each laser. $\hat{\varepsilon}_{0}$, $\hat{\varepsilon}_{+}$ and $\hat{\varepsilon}_{-}$ are $\pi$, $\sigma^{+}$ and $\sigma^{-}$ polarization vectors, respectively. Using second order perturbation theory\cite{cohen1998atom,wright2008raman,supplemental}, the Raman coupling Hamiltonian can be made into the desired form. In our calculations, we apply $\hbar\omega=20E_r$ with Raman laser wavelengths close to $791\textrm{nm}$ and $806\textrm{nm}$, which are achievable in the laboratory. These parameters result in a total laser power of around $0.8\textrm{W}$ with Gaussian laser beams with beam waists of $200 \upmu$m, and a scattering limited lifetime of $\tau_{sc}\approx133\textrm{ms}$, which is enough for experimental applications.

To load the system into the ground state of the TMDOL, we can start from a cigar-shaped Bose--Einstein condensate with the desired atom number in an optical dipole trap with the long axis of the trap aligned with $x$-axis in the lab frame. Then, we adiabatically ramp up the two-dimensional optical lattice along the $y$ and $z$-axes to avoid excitation to higher bands. Next, applying the shortcut adiabatic loading technique \cite{torrontegui2011fast,corgier2018fast,guery2019shortcuts,ding2020smooth} to load the Bose gas into the ground state of a moving 2D optical lattice. Due to the relative motion between the lattice along the $z$-axis and the lab frame, the atoms will carry a transverse momentum with expectation value $\langle p_z\rangle=-\hbar\bar{k}_z$ in the moving lattice frame. Then, adiabatically turn on the Raman lasers and the system will have been loaded into the ground state of the TMDOL with a certain $\langle p_z\rangle$. To make observations, one can suddenly shut off all lasers at the end of the periodic driving cycle when the micromotion operator, $U$, becomes the identity operator so that the system is automatically projected from the Floquet basis to the Zeeman sublevel basis\cite{chen20202}. Then, a time-of-flight (TOF) image can be taken with the imaging beam aligned along the $z$-axis. The interference pattern contrast, $\mathcal{I}$, can be measured from the TOF image \cite{gerbier2005interference,soltan2011multi}, which signals SF and MI phases.

%%%%%%%%%%%%%%%%%%%%%%%%%%%%%%%%%
\lettersection{Conclusion}

To conclude, we proposed an experimentally feasible method to create an effective 1D TMD optical lattice induced by an artificial gauge potential in a periodically driven system with Raman couplings. We explored the many-body physics of the system induced by a periodic artificial gauge potential by constructing the TMD effective 1D Bose-Hubbard model with a deep 2D transverse confinement. Using the strong coupling expansion and QMC methods, we calculated the phase diagram of an effective 1D TMDOL, and showed that the SF-MI transition can be tuned by changing the average transverse momentum of the Bose gas. Then, we proposed a possible experimental implementation that achieves the desired periodically driven Hamiltonian and includes loading and measurement procedures.

Our work provides a new platform for studying the many-body physics of a cold atom system with an artificial gauge potential. For instance, the TMDOL makes it possible to create two spatially overlapping BECs of the same species in different quantum phases, which gives rise to opportunities to study if the two BECs, one SF and one MI, will exchange their phases during a collision in which they exchange their momenta. Additionally, more complicated TMDOL potentials may be considered by engineering the transverse momentum distribution of the Bose gas and introducing Raman lasers along other directions to achieve a 2D TMDOL. The transverse momentum in a TMDOL can also be regarded as a synthetic dimension, which is analogous to spin or harmonic trap eigenstates in previous work \cite{celi2014synthetic,taddia2017topological,price2017synthetic}. Finally, for fermionic systems, one could also implement a TMDOL with appropriate Raman couplings, and such a system provides a broad playground for studying TMD spin dynamics and many-body phases.

\begin{acknowledgments}
We thank Thierry Giamarchi for insightful discussions. This work is supported by NSF grant PHY 1708008, NASA/JPL RSA 1656126 and the Swiss National Science Foundation under Division II. Numerical calculations make use of the ALPS scheduler library and statistical analysis tools~\cite{troyer1998,ALPS2007,ALPS2011}.
\end{acknowledgments}

\bibliography{reference}

\begin{thebibliography}{56}%
\makeatletter
\providecommand \@ifxundefined [1]{%
 \@ifx{#1\undefined}
}%
\providecommand \@ifnum [1]{%
 \ifnum #1\expandafter \@firstoftwo
 \else \expandafter \@secondoftwo
 \fi
}%
\providecommand \@ifx [1]{%
 \ifx #1\expandafter \@firstoftwo
 \else \expandafter \@secondoftwo
 \fi
}%
\providecommand \natexlab [1]{#1}%
\providecommand \enquote  [1]{``#1''}%
\providecommand \bibnamefont  [1]{#1}%
\providecommand \bibfnamefont [1]{#1}%
\providecommand \citenamefont [1]{#1}%
\providecommand \href@noop [0]{\@secondoftwo}%
\providecommand \href [0]{\begingroup \@sanitize@url \@href}%
\providecommand \@href[1]{\@@startlink{#1}\@@href}%
\providecommand \@@href[1]{\endgroup#1\@@endlink}%
\providecommand \@sanitize@url [0]{\catcode `\\12\catcode `\$12\catcode
  `\&12\catcode `\#12\catcode `\^12\catcode `\_12\catcode `\%12\relax}%
\providecommand \@@startlink[1]{}%
\providecommand \@@endlink[0]{}%
\providecommand \url  [0]{\begingroup\@sanitize@url \@url }%
\providecommand \@url [1]{\endgroup\@href {#1}{\urlprefix }}%
\providecommand \urlprefix  [0]{URL }%
\providecommand \Eprint [0]{\href }%
\providecommand \doibase [0]{http://dx.doi.org/}%
\providecommand \selectlanguage [0]{\@gobble}%
\providecommand \bibinfo  [0]{\@secondoftwo}%
\providecommand \bibfield  [0]{\@secondoftwo}%
\providecommand \translation [1]{[#1]}%
\providecommand \BibitemOpen [0]{}%
\providecommand \bibitemStop [0]{}%
\providecommand \bibitemNoStop [0]{.\EOS\space}%
\providecommand \EOS [0]{\spacefactor3000\relax}%
\providecommand \BibitemShut  [1]{\csname bibitem#1\endcsname}%
\let\auto@bib@innerbib\@empty
%</preamble>
\bibitem [{\citenamefont {Sugawa}\ \emph {et~al.}(2018)\citenamefont {Sugawa},
  \citenamefont {Salces-Carcoba}, \citenamefont {Perry}, \citenamefont {Yue},\
  and\ \citenamefont {Spielman}}]{sugawa2018second}%
  \BibitemOpen
  \bibfield  {author} {\bibinfo {author} {\bibfnamefont {S.}~\bibnamefont
  {Sugawa}}, \bibinfo {author} {\bibfnamefont {F.}~\bibnamefont
  {Salces-Carcoba}}, \bibinfo {author} {\bibfnamefont {A.~R.}\ \bibnamefont
  {Perry}}, \bibinfo {author} {\bibfnamefont {Y.}~\bibnamefont {Yue}}, \ and\
  \bibinfo {author} {\bibfnamefont {I.~B.}\ \bibnamefont {Spielman}},\
  }\href@noop {} {\bibfield  {journal} {\bibinfo  {journal} {Science}\ }\textbf
  {\bibinfo {volume} {360}},\ \bibinfo {pages} {1429--1434} (\bibinfo {year}
  {2018})}\BibitemShut {NoStop}%
\bibitem [{\citenamefont {Zhu}\ \emph {et~al.}(2006)\citenamefont {Zhu},
  \citenamefont {Fu}, \citenamefont {Wu}, \citenamefont {Zhang},\ and\
  \citenamefont {Duan}}]{zhu2006spin}%
  \BibitemOpen
  \bibfield  {author} {\bibinfo {author} {\bibfnamefont {S.-L.}\ \bibnamefont
  {Zhu}}, \bibinfo {author} {\bibfnamefont {H.}~\bibnamefont {Fu}}, \bibinfo
  {author} {\bibfnamefont {C.-J.}\ \bibnamefont {Wu}}, \bibinfo {author}
  {\bibfnamefont {S.-C.}\ \bibnamefont {Zhang}}, \ and\ \bibinfo {author}
  {\bibfnamefont {L.-M.}\ \bibnamefont {Duan}},\ }\href@noop {} {\bibfield
  {journal} {\bibinfo  {journal} {\Jprl}\ }\textbf {\bibinfo {volume} {97}},\
  \bibinfo {pages} {240401} (\bibinfo {year} {2006})}\BibitemShut {NoStop}%
\bibitem [{\citenamefont {Liu}\ \emph {et~al.}(2007)\citenamefont {Liu},
  \citenamefont {Liu}, \citenamefont {Kwek},\ and\ \citenamefont
  {Oh}}]{liu2007optically}%
  \BibitemOpen
  \bibfield  {author} {\bibinfo {author} {\bibfnamefont {X.-J.}\ \bibnamefont
  {Liu}}, \bibinfo {author} {\bibfnamefont {X.}~\bibnamefont {Liu}}, \bibinfo
  {author} {\bibfnamefont {L.~C.}\ \bibnamefont {Kwek}}, \ and\ \bibinfo
  {author} {\bibfnamefont {C.~H.}\ \bibnamefont {Oh}},\ }\href@noop {}
  {\bibfield  {journal} {\bibinfo  {journal} {\Jprl}\ }\textbf {\bibinfo
  {volume} {98}},\ \bibinfo {pages} {026602} (\bibinfo {year}
  {2007})}\BibitemShut {NoStop}%
\bibitem [{\citenamefont {LeBlanc}\ \emph {et~al.}(2012)\citenamefont
  {LeBlanc}, \citenamefont {Jim{\'e}nez-Garc{\'\i}a}, \citenamefont {Williams},
  \citenamefont {Beeler}, \citenamefont {Perry}, \citenamefont {Phillips},\
  and\ \citenamefont {Spielman}}]{leblanc2012observation}%
  \BibitemOpen
  \bibfield  {author} {\bibinfo {author} {\bibfnamefont {L.~J.}\ \bibnamefont
  {LeBlanc}}, \bibinfo {author} {\bibfnamefont {K.}~\bibnamefont
  {Jim{\'e}nez-Garc{\'\i}a}}, \bibinfo {author} {\bibfnamefont {R.~A.}\
  \bibnamefont {Williams}}, \bibinfo {author} {\bibfnamefont {M.~C.}\
  \bibnamefont {Beeler}}, \bibinfo {author} {\bibfnamefont {A.~R.}\
  \bibnamefont {Perry}}, \bibinfo {author} {\bibfnamefont {W.~D.}\ \bibnamefont
  {Phillips}}, \ and\ \bibinfo {author} {\bibfnamefont {I.~B.}\ \bibnamefont
  {Spielman}},\ }\href@noop {} {\bibfield  {journal} {\bibinfo  {journal}
  {Proceedings of the National Academy of Sciences}\ }\textbf {\bibinfo
  {volume} {109}},\ \bibinfo {pages} {10811--10814} (\bibinfo {year}
  {2012})}\BibitemShut {NoStop}%
\bibitem [{\citenamefont {Beeler}\ \emph {et~al.}(2013)\citenamefont {Beeler},
  \citenamefont {Williams}, \citenamefont {Jimenez-Garcia}, \citenamefont
  {LeBlanc}, \citenamefont {Perry},\ and\ \citenamefont
  {Spielman}}]{beeler2013spin}%
  \BibitemOpen
  \bibfield  {author} {\bibinfo {author} {\bibfnamefont {M.~C.}\ \bibnamefont
  {Beeler}}, \bibinfo {author} {\bibfnamefont {R.~A.}\ \bibnamefont
  {Williams}}, \bibinfo {author} {\bibfnamefont {K.}~\bibnamefont
  {Jimenez-Garcia}}, \bibinfo {author} {\bibfnamefont {L.~J.}\ \bibnamefont
  {LeBlanc}}, \bibinfo {author} {\bibfnamefont {A.~R.}\ \bibnamefont {Perry}},
  \ and\ \bibinfo {author} {\bibfnamefont {I.~B.}\ \bibnamefont {Spielman}},\
  }\href@noop {} {\bibfield  {journal} {\bibinfo  {journal} {Nature}\ }\textbf
  {\bibinfo {volume} {498}},\ \bibinfo {pages} {201--204} (\bibinfo {year}
  {2013})}\BibitemShut {NoStop}%
\bibitem [{\citenamefont {Dalibard}\ \emph {et~al.}(2011)\citenamefont
  {Dalibard}, \citenamefont {Gerbier}, \citenamefont {Juzeli{\=u}nas},\ and\
  \citenamefont {{\"O}hberg}}]{dalibard2011colloquium}%
  \BibitemOpen
  \bibfield  {author} {\bibinfo {author} {\bibfnamefont {J.}~\bibnamefont
  {Dalibard}}, \bibinfo {author} {\bibfnamefont {F.}~\bibnamefont {Gerbier}},
  \bibinfo {author} {\bibfnamefont {G.}~\bibnamefont {Juzeli{\=u}nas}}, \ and\
  \bibinfo {author} {\bibfnamefont {P.}~\bibnamefont {{\"O}hberg}},\
  }\href@noop {} {\bibfield  {journal} {\bibinfo  {journal} {Rev. Mod. Phys.}\
  }\textbf {\bibinfo {volume} {83}},\ \bibinfo {pages} {1523} (\bibinfo {year}
  {2011})}\BibitemShut {NoStop}%
\bibitem [{\citenamefont {Goldman}\ \emph {et~al.}(2014)\citenamefont
  {Goldman}, \citenamefont {Juzeli{\=u}nas}, \citenamefont {{\"O}hberg},\ and\
  \citenamefont {Spielman}}]{goldman2014light}%
  \BibitemOpen
  \bibfield  {author} {\bibinfo {author} {\bibfnamefont {N.}~\bibnamefont
  {Goldman}}, \bibinfo {author} {\bibfnamefont {G.}~\bibnamefont
  {Juzeli{\=u}nas}}, \bibinfo {author} {\bibfnamefont {P.}~\bibnamefont
  {{\"O}hberg}}, \ and\ \bibinfo {author} {\bibfnamefont {I.~B.}\ \bibnamefont
  {Spielman}},\ }\href@noop {} {\bibfield  {journal} {\bibinfo  {journal}
  {Reports on Progress in Physics}\ }\textbf {\bibinfo {volume} {77}},\
  \bibinfo {pages} {126401} (\bibinfo {year} {2014})}\BibitemShut {NoStop}%
\bibitem [{\citenamefont {Zhai}(2015)}]{zhai2015degenerate}%
  \BibitemOpen
  \bibfield  {author} {\bibinfo {author} {\bibfnamefont {H.}~\bibnamefont
  {Zhai}},\ }\href@noop {} {\bibfield  {journal} {\bibinfo  {journal} {Reports
  on Progress in Physics}\ }\textbf {\bibinfo {volume} {78}},\ \bibinfo {pages}
  {026001} (\bibinfo {year} {2015})}\BibitemShut {NoStop}%
\bibitem [{\citenamefont {Zhang}\ and\ \citenamefont
  {Liu}(2018)}]{zhang2018spin}%
  \BibitemOpen
  \bibfield  {author} {\bibinfo {author} {\bibfnamefont {L.}~\bibnamefont
  {Zhang}}\ and\ \bibinfo {author} {\bibfnamefont {X.-J.}\ \bibnamefont
  {Liu}},\ }in\ \href@noop {} {\emph {\bibinfo {booktitle} {Synthetic
  Spin-Orbit Coupling in Cold Atoms}}}\ (\bibinfo  {publisher} {World
  Scientific},\ \bibinfo {year} {2018})\ pp.\ \bibinfo {pages}
  {1--87}\BibitemShut {NoStop}%
\bibitem [{\citenamefont {Huang}\ \emph {et~al.}(2016)\citenamefont {Huang},
  \citenamefont {Meng}, \citenamefont {Wang}, \citenamefont {Peng},
  \citenamefont {Zhang}, \citenamefont {Chen}, \citenamefont {Li},
  \citenamefont {Zhou},\ and\ \citenamefont {Zhang}}]{huang2016experimental}%
  \BibitemOpen
  \bibfield  {author} {\bibinfo {author} {\bibfnamefont {L.}~\bibnamefont
  {Huang}}, \bibinfo {author} {\bibfnamefont {Z.}~\bibnamefont {Meng}},
  \bibinfo {author} {\bibfnamefont {P.}~\bibnamefont {Wang}}, \bibinfo {author}
  {\bibfnamefont {P.}~\bibnamefont {Peng}}, \bibinfo {author} {\bibfnamefont
  {S.-L.}\ \bibnamefont {Zhang}}, \bibinfo {author} {\bibfnamefont
  {L.}~\bibnamefont {Chen}}, \bibinfo {author} {\bibfnamefont {D.}~\bibnamefont
  {Li}}, \bibinfo {author} {\bibfnamefont {Q.}~\bibnamefont {Zhou}}, \ and\
  \bibinfo {author} {\bibfnamefont {J.}~\bibnamefont {Zhang}},\ }\href@noop {}
  {\bibfield  {journal} {\bibinfo  {journal} {Nature Physics}\ }\textbf
  {\bibinfo {volume} {12}},\ \bibinfo {pages} {540--544} (\bibinfo {year}
  {2016})}\BibitemShut {NoStop}%
\bibitem [{\citenamefont {Hamner}\ \emph {et~al.}(2015)\citenamefont {Hamner},
  \citenamefont {Zhang}, \citenamefont {Khamehchi}, \citenamefont {Davis},\
  and\ \citenamefont {Engels}}]{hamner2015spin}%
  \BibitemOpen
  \bibfield  {author} {\bibinfo {author} {\bibfnamefont {C.}~\bibnamefont
  {Hamner}}, \bibinfo {author} {\bibfnamefont {Y.}~\bibnamefont {Zhang}},
  \bibinfo {author} {\bibfnamefont {M.~A.}\ \bibnamefont {Khamehchi}}, \bibinfo
  {author} {\bibfnamefont {M.~J.}\ \bibnamefont {Davis}}, \ and\ \bibinfo
  {author} {\bibfnamefont {P.}~\bibnamefont {Engels}},\ }\href@noop {}
  {\bibfield  {journal} {\bibinfo  {journal} {\Jprl}\ }\textbf {\bibinfo
  {volume} {114}},\ \bibinfo {pages} {070401} (\bibinfo {year}
  {2015})}\BibitemShut {NoStop}%
\bibitem [{\citenamefont {Lin}\ \emph {et~al.}(2011)\citenamefont {Lin},
  \citenamefont {Jim{\'e}nez-Garc{\'\i}a},\ and\ \citenamefont
  {Spielman}}]{lin2011spin}%
  \BibitemOpen
  \bibfield  {author} {\bibinfo {author} {\bibfnamefont {Y.-J.}\ \bibnamefont
  {Lin}}, \bibinfo {author} {\bibfnamefont {K.}~\bibnamefont
  {Jim{\'e}nez-Garc{\'\i}a}}, \ and\ \bibinfo {author} {\bibfnamefont {I.~B.}\
  \bibnamefont {Spielman}},\ }\href@noop {} {\bibfield  {journal} {\bibinfo
  {journal} {Nature}\ }\textbf {\bibinfo {volume} {471}},\ \bibinfo {pages}
  {83--86} (\bibinfo {year} {2011})}\BibitemShut {NoStop}%
\bibitem [{\citenamefont {Liu}\ \emph {et~al.}(2013)\citenamefont {Liu},
  \citenamefont {Liu},\ and\ \citenamefont {Cheng}}]{liu2013manipulating}%
  \BibitemOpen
  \bibfield  {author} {\bibinfo {author} {\bibfnamefont {X.-J.}\ \bibnamefont
  {Liu}}, \bibinfo {author} {\bibfnamefont {Z.-X.}\ \bibnamefont {Liu}}, \ and\
  \bibinfo {author} {\bibfnamefont {M.}~\bibnamefont {Cheng}},\ }\href@noop {}
  {\bibfield  {journal} {\bibinfo  {journal} {\Jprl}\ }\textbf {\bibinfo
  {volume} {110}},\ \bibinfo {pages} {076401} (\bibinfo {year}
  {2013})}\BibitemShut {NoStop}%
\bibitem [{\citenamefont {Wu}\ \emph {et~al.}(2016)\citenamefont {Wu},
  \citenamefont {Zhang}, \citenamefont {Sun}, \citenamefont {Xu}, \citenamefont
  {Wang}, \citenamefont {Ji}, \citenamefont {Deng}, \citenamefont {Chen},
  \citenamefont {Liu},\ and\ \citenamefont {Pan}}]{wu2016realization}%
  \BibitemOpen
  \bibfield  {author} {\bibinfo {author} {\bibfnamefont {Z.}~\bibnamefont
  {Wu}}, \bibinfo {author} {\bibfnamefont {L.}~\bibnamefont {Zhang}}, \bibinfo
  {author} {\bibfnamefont {W.}~\bibnamefont {Sun}}, \bibinfo {author}
  {\bibfnamefont {X.-T.}\ \bibnamefont {Xu}}, \bibinfo {author} {\bibfnamefont
  {B.-Z.}\ \bibnamefont {Wang}}, \bibinfo {author} {\bibfnamefont {S.-C.}\
  \bibnamefont {Ji}}, \bibinfo {author} {\bibfnamefont {Y.}~\bibnamefont
  {Deng}}, \bibinfo {author} {\bibfnamefont {S.}~\bibnamefont {Chen}}, \bibinfo
  {author} {\bibfnamefont {X.-J.}\ \bibnamefont {Liu}}, \ and\ \bibinfo
  {author} {\bibfnamefont {J.-W.}\ \bibnamefont {Pan}},\ }\href@noop {}
  {\bibfield  {journal} {\bibinfo  {journal} {Science}\ }\textbf {\bibinfo
  {volume} {354}},\ \bibinfo {pages} {83--88} (\bibinfo {year}
  {2016})}\BibitemShut {NoStop}%
\bibitem [{\citenamefont {et~al.}(2021)}]{wang2021realization}%
  \BibitemOpen
  \bibfield  {author} {\bibinfo {author} {\bibfnamefont {Z.-Y.~Wang}\
  \bibnamefont {et~al.}},\ }\href@noop {} {\bibfield  {journal} {\bibinfo
  {journal} {Science}\ }\textbf {\bibinfo {volume} {372}},\ \bibinfo {pages}
  {271--276} (\bibinfo {year} {2021})}\BibitemShut {NoStop}%
\bibitem [{\citenamefont {Bloch}\ \emph {et~al.}(2008)\citenamefont {Bloch},
  \citenamefont {Dalibard},\ and\ \citenamefont {Zwerger}}]{bloch2008many}%
  \BibitemOpen
  \bibfield  {author} {\bibinfo {author} {\bibfnamefont {I.}~\bibnamefont
  {Bloch}}, \bibinfo {author} {\bibfnamefont {J.}~\bibnamefont {Dalibard}}, \
  and\ \bibinfo {author} {\bibfnamefont {W.}~\bibnamefont {Zwerger}},\
  }\href@noop {} {\bibfield  {journal} {\bibinfo  {journal} {Rev. Mod. Phys.}\
  }\textbf {\bibinfo {volume} {80}},\ \bibinfo {pages} {885} (\bibinfo {year}
  {2008})}\BibitemShut {NoStop}%
\bibitem [{\citenamefont {Jaksch}\ \emph {et~al.}(1998)\citenamefont {Jaksch},
  \citenamefont {Bruder}, \citenamefont {Cirac}, \citenamefont {Gardiner},\
  and\ \citenamefont {Zoller}}]{jaksch1998cold}%
  \BibitemOpen
  \bibfield  {author} {\bibinfo {author} {\bibfnamefont {D.}~\bibnamefont
  {Jaksch}}, \bibinfo {author} {\bibfnamefont {C.}~\bibnamefont {Bruder}},
  \bibinfo {author} {\bibfnamefont {J.~I.}\ \bibnamefont {Cirac}}, \bibinfo
  {author} {\bibfnamefont {C.~W.}\ \bibnamefont {Gardiner}}, \ and\ \bibinfo
  {author} {\bibfnamefont {P.}~\bibnamefont {Zoller}},\ }\href@noop {}
  {\bibfield  {journal} {\bibinfo  {journal} {\Jprl}\ }\textbf {\bibinfo
  {volume} {81}},\ \bibinfo {pages} {3108} (\bibinfo {year}
  {1998})}\BibitemShut {NoStop}%
\bibitem [{\citenamefont {Bo\'eris}\ \emph {et~al.}(2016)\citenamefont
  {Bo\'eris}, \citenamefont {Gori}, \citenamefont {Hoogerland}, \citenamefont
  {Kumar}, \citenamefont {Lucioni}, \citenamefont {Tanzi}, \citenamefont
  {Inguscio}, \citenamefont {Giamarchi}, \citenamefont {D'Errico},
  \citenamefont {Carleo}, \citenamefont {Modugno},\ and\ \citenamefont
  {Sanchez-Palencia}}]{boeris2016}%
  \BibitemOpen
  \bibfield  {author} {\bibinfo {author} {\bibfnamefont {G.}~\bibnamefont
  {Bo\'eris}}, \bibinfo {author} {\bibfnamefont {L.}~\bibnamefont {Gori}},
  \bibinfo {author} {\bibfnamefont {M.~D.}\ \bibnamefont {Hoogerland}},
  \bibinfo {author} {\bibfnamefont {A.}~\bibnamefont {Kumar}}, \bibinfo
  {author} {\bibfnamefont {E.}~\bibnamefont {Lucioni}}, \bibinfo {author}
  {\bibfnamefont {L.}~\bibnamefont {Tanzi}}, \bibinfo {author} {\bibfnamefont
  {M.}~\bibnamefont {Inguscio}}, \bibinfo {author} {\bibfnamefont
  {T.}~\bibnamefont {Giamarchi}}, \bibinfo {author} {\bibfnamefont
  {C.}~\bibnamefont {D'Errico}}, \bibinfo {author} {\bibfnamefont
  {G.}~\bibnamefont {Carleo}}, \bibinfo {author} {\bibfnamefont
  {G.}~\bibnamefont {Modugno}}, \ and\ \bibinfo {author} {\bibfnamefont
  {L.}~\bibnamefont {Sanchez-Palencia}},\ }\href@noop {} {\bibfield  {journal}
  {\bibinfo  {journal} {\Jpra}\ }\textbf {\bibinfo {volume} {93}},\ \bibinfo
  {pages} {011601(R)} (\bibinfo {year} {2016})}\BibitemShut {NoStop}%
\bibitem [{\citenamefont {Greiner}\ \emph {et~al.}(2002)\citenamefont
  {Greiner}, \citenamefont {Mandel}, \citenamefont {Esslinger}, \citenamefont
  {H{\"a}nsch},\ and\ \citenamefont {Bloch}}]{greiner2002quantum}%
  \BibitemOpen
  \bibfield  {author} {\bibinfo {author} {\bibfnamefont {M.}~\bibnamefont
  {Greiner}}, \bibinfo {author} {\bibfnamefont {O.}~\bibnamefont {Mandel}},
  \bibinfo {author} {\bibfnamefont {T.}~\bibnamefont {Esslinger}}, \bibinfo
  {author} {\bibfnamefont {T.~W.}\ \bibnamefont {H{\"a}nsch}}, \ and\ \bibinfo
  {author} {\bibfnamefont {I.}~\bibnamefont {Bloch}},\ }\href@noop {}
  {\bibfield  {journal} {\bibinfo  {journal} {Nature}\ }\textbf {\bibinfo
  {volume} {415}},\ \bibinfo {pages} {39--44} (\bibinfo {year}
  {2002})}\BibitemShut {NoStop}%
\bibitem [{\citenamefont {Haller}\ \emph {et~al.}(2010)\citenamefont {Haller},
  \citenamefont {Hart}, \citenamefont {Mark}, \citenamefont {Danzl},
  \citenamefont {Reichs\"ollner}, \citenamefont {Gustavsson}, \citenamefont
  {Dalmonte}, \citenamefont {Pupillo},\ and\ \citenamefont
  {N\"agerl}}]{haller2010}%
  \BibitemOpen
  \bibfield  {author} {\bibinfo {author} {\bibfnamefont {E.}~\bibnamefont
  {Haller}}, \bibinfo {author} {\bibfnamefont {R.}~\bibnamefont {Hart}},
  \bibinfo {author} {\bibfnamefont {M.~J.}\ \bibnamefont {Mark}}, \bibinfo
  {author} {\bibfnamefont {J.~G.}\ \bibnamefont {Danzl}}, \bibinfo {author}
  {\bibfnamefont {L.}~\bibnamefont {Reichs\"ollner}}, \bibinfo {author}
  {\bibfnamefont {M.}~\bibnamefont {Gustavsson}}, \bibinfo {author}
  {\bibfnamefont {M.}~\bibnamefont {Dalmonte}}, \bibinfo {author}
  {\bibfnamefont {G.}~\bibnamefont {Pupillo}}, \ and\ \bibinfo {author}
  {\bibfnamefont {H.-C.}\ \bibnamefont {N\"agerl}},\ }\href@noop {} {\bibfield
  {journal} {\bibinfo  {journal} {Nature}\ }\textbf {\bibinfo {volume} {466}},\
  \bibinfo {pages} {597--600} (\bibinfo {year} {2010})}\BibitemShut {NoStop}%
\bibitem [{\citenamefont {Soltan-Panahi}\ \emph {et~al.}(2011)\citenamefont
  {Soltan-Panahi}, \citenamefont {Struck}, \citenamefont {Hauke}, \citenamefont
  {Bick}, \citenamefont {Plenkers}, \citenamefont {Meineke}, \citenamefont
  {Becker}, \citenamefont {Windpassinger}, \citenamefont {Lewenstein},\ and\
  \citenamefont {Sengstock}}]{soltan2011multi}%
  \BibitemOpen
  \bibfield  {author} {\bibinfo {author} {\bibfnamefont {P.}~\bibnamefont
  {Soltan-Panahi}}, \bibinfo {author} {\bibfnamefont {J.}~\bibnamefont
  {Struck}}, \bibinfo {author} {\bibfnamefont {P.}~\bibnamefont {Hauke}},
  \bibinfo {author} {\bibfnamefont {A.}~\bibnamefont {Bick}}, \bibinfo {author}
  {\bibfnamefont {W.}~\bibnamefont {Plenkers}}, \bibinfo {author}
  {\bibfnamefont {G.}~\bibnamefont {Meineke}}, \bibinfo {author} {\bibfnamefont
  {C.}~\bibnamefont {Becker}}, \bibinfo {author} {\bibfnamefont
  {P.}~\bibnamefont {Windpassinger}}, \bibinfo {author} {\bibfnamefont
  {M.}~\bibnamefont {Lewenstein}}, \ and\ \bibinfo {author} {\bibfnamefont
  {K.}~\bibnamefont {Sengstock}},\ }\href@noop {} {\bibfield  {journal}
  {\bibinfo  {journal} {Nature Physics}\ }\textbf {\bibinfo {volume} {7}},\
  \bibinfo {pages} {434--440} (\bibinfo {year} {2011})}\BibitemShut {NoStop}%
\bibitem [{\citenamefont {Cai}\ \emph {et~al.}(2012)\citenamefont {Cai},
  \citenamefont {Zhou},\ and\ \citenamefont {Wu}}]{cai2012magnetic}%
  \BibitemOpen
  \bibfield  {author} {\bibinfo {author} {\bibfnamefont {Z.}~\bibnamefont
  {Cai}}, \bibinfo {author} {\bibfnamefont {X.}~\bibnamefont {Zhou}}, \ and\
  \bibinfo {author} {\bibfnamefont {C.}~\bibnamefont {Wu}},\ }\href@noop {}
  {\bibfield  {journal} {\bibinfo  {journal} {\Jpra}\ }\textbf {\bibinfo
  {volume} {85}},\ \bibinfo {pages} {061605} (\bibinfo {year}
  {2012})}\BibitemShut {NoStop}%
\bibitem [{\citenamefont {Cole}\ \emph {et~al.}(2012)\citenamefont {Cole},
  \citenamefont {Zhang}, \citenamefont {Paramekanti},\ and\ \citenamefont
  {Trivedi}}]{cole2012bose}%
  \BibitemOpen
  \bibfield  {author} {\bibinfo {author} {\bibfnamefont {W.~S.}\ \bibnamefont
  {Cole}}, \bibinfo {author} {\bibfnamefont {S.}~\bibnamefont {Zhang}},
  \bibinfo {author} {\bibfnamefont {A.}~\bibnamefont {Paramekanti}}, \ and\
  \bibinfo {author} {\bibfnamefont {N.}~\bibnamefont {Trivedi}},\ }\href@noop
  {} {\bibfield  {journal} {\bibinfo  {journal} {\Jprl}\ }\textbf {\bibinfo
  {volume} {109}},\ \bibinfo {pages} {085302} (\bibinfo {year}
  {2012})}\BibitemShut {NoStop}%
\bibitem [{\citenamefont {Ra{\v{c}}kauskas}\ \emph {et~al.}(2019)\citenamefont
  {Ra{\v{c}}kauskas}, \citenamefont {Novi{\v{c}}enko}, \citenamefont {Pu},\
  and\ \citenamefont {Juzeli{\=u}nas}}]{ravckauskas2019non}%
  \BibitemOpen
  \bibfield  {author} {\bibinfo {author} {\bibfnamefont {P.}~\bibnamefont
  {Ra{\v{c}}kauskas}}, \bibinfo {author} {\bibfnamefont {V.}~\bibnamefont
  {Novi{\v{c}}enko}}, \bibinfo {author} {\bibfnamefont {H.}~\bibnamefont {Pu}},
  \ and\ \bibinfo {author} {\bibfnamefont {G.}~\bibnamefont {Juzeli{\=u}nas}},\
  }\href@noop {} {\bibfield  {journal} {\bibinfo  {journal} {\Jpra}\ }\textbf
  {\bibinfo {volume} {100}},\ \bibinfo {pages} {063616} (\bibinfo {year}
  {2019})}\BibitemShut {NoStop}%
\bibitem [{\citenamefont {Novi{\v{c}}enko}\ and\ \citenamefont
  {Juzeli{\=u}nas}(2019)}]{novivcenko2019non}%
  \BibitemOpen
  \bibfield  {author} {\bibinfo {author} {\bibfnamefont {V.}~\bibnamefont
  {Novi{\v{c}}enko}}\ and\ \bibinfo {author} {\bibfnamefont {G.}~\bibnamefont
  {Juzeli{\=u}nas}},\ }\href@noop {} {\bibfield  {journal} {\bibinfo  {journal}
  {\Jpra}\ }\textbf {\bibinfo {volume} {100}},\ \bibinfo {pages} {012127}
  (\bibinfo {year} {2019})}\BibitemShut {NoStop}%
\bibitem [{\citenamefont {Chen}\ \emph {et~al.}(2020)\citenamefont {Chen},
  \citenamefont {Murphree},\ and\ \citenamefont {Bigelow}}]{chen20202}%
  \BibitemOpen
  \bibfield  {author} {\bibinfo {author} {\bibfnamefont {Z.}~\bibnamefont
  {Chen}}, \bibinfo {author} {\bibfnamefont {J.~D.}\ \bibnamefont {Murphree}},
  \ and\ \bibinfo {author} {\bibfnamefont {N.~P.}\ \bibnamefont {Bigelow}},\
  }\href@noop {} {\bibfield  {journal} {\bibinfo  {journal} {\Jpra}\ }\textbf
  {\bibinfo {volume} {101}},\ \bibinfo {pages} {013606} (\bibinfo {year}
  {2020})}\BibitemShut {NoStop}%
\bibitem [{sup()}]{supplemental}%
  \BibitemOpen
  \href@noop {} {}\bibinfo {note} {See supplemental material. We discuss
  calculation of Floquet Hamiltonian and the adiabatic conditions, the single
  particle ground state of the TMDOL, the $\bar{n}=2$ Mott lobe, and the
  detailed construction of the periodically driven Hamiltonian.}\BibitemShut
  {Stop}%
\bibitem [{\citenamefont {St{\"o}ferle}\ \emph {et~al.}(2004)\citenamefont
  {St{\"o}ferle}, \citenamefont {Moritz}, \citenamefont {Schori}, \citenamefont
  {K{\"o}hl},\ and\ \citenamefont {Esslinger}}]{stoferle2004transition}%
  \BibitemOpen
  \bibfield  {author} {\bibinfo {author} {\bibfnamefont {T.}~\bibnamefont
  {St{\"o}ferle}}, \bibinfo {author} {\bibfnamefont {H.}~\bibnamefont
  {Moritz}}, \bibinfo {author} {\bibfnamefont {C.}~\bibnamefont {Schori}},
  \bibinfo {author} {\bibfnamefont {M.}~\bibnamefont {K{\"o}hl}}, \ and\
  \bibinfo {author} {\bibfnamefont {T.}~\bibnamefont {Esslinger}},\ }\href@noop
  {} {\bibfield  {journal} {\bibinfo  {journal} {\Jprl}\ }\textbf {\bibinfo
  {volume} {92}},\ \bibinfo {pages} {130403} (\bibinfo {year}
  {2004})}\BibitemShut {NoStop}%
\bibitem [{\citenamefont {Anisimovas}\ \emph {et~al.}(2016)\citenamefont
  {Anisimovas}, \citenamefont {Ra{\v{c}}i{\=u}nas}, \citenamefont
  {Str{\"a}ter}, \citenamefont {Eckardt}, \citenamefont {Spielman},\ and\
  \citenamefont {Juzeli{\=u}nas}}]{anisimovas2016semisynthetic}%
  \BibitemOpen
  \bibfield  {author} {\bibinfo {author} {\bibfnamefont {E.}~\bibnamefont
  {Anisimovas}}, \bibinfo {author} {\bibfnamefont {M.}~\bibnamefont
  {Ra{\v{c}}i{\=u}nas}}, \bibinfo {author} {\bibfnamefont {C.}~\bibnamefont
  {Str{\"a}ter}}, \bibinfo {author} {\bibfnamefont {A.}~\bibnamefont
  {Eckardt}}, \bibinfo {author} {\bibfnamefont {I.~B.}\ \bibnamefont
  {Spielman}}, \ and\ \bibinfo {author} {\bibfnamefont {G.}~\bibnamefont
  {Juzeli{\=u}nas}},\ }\href@noop {} {\bibfield  {journal} {\bibinfo  {journal}
  {\Jpra}\ }\textbf {\bibinfo {volume} {94}},\ \bibinfo {pages} {063632}
  (\bibinfo {year} {2016})}\BibitemShut {NoStop}%
\bibitem [{\citenamefont {Barbiero}\ and\ \citenamefont
  {Salasnich}(2014)}]{barbiero2014quantum}%
  \BibitemOpen
  \bibfield  {author} {\bibinfo {author} {\bibfnamefont {L.}~\bibnamefont
  {Barbiero}}\ and\ \bibinfo {author} {\bibfnamefont {L.}~\bibnamefont
  {Salasnich}},\ }\href@noop {} {\bibfield  {journal} {\bibinfo  {journal}
  {\Jpra}\ }\textbf {\bibinfo {volume} {89}},\ \bibinfo {pages} {063605}
  (\bibinfo {year} {2014})}\BibitemShut {NoStop}%
\bibitem [{\citenamefont {K{\"u}hner}\ and\ \citenamefont
  {Monien}(1998)}]{kuhner1998phases}%
  \BibitemOpen
  \bibfield  {author} {\bibinfo {author} {\bibfnamefont {T.~D.}\ \bibnamefont
  {K{\"u}hner}}\ and\ \bibinfo {author} {\bibfnamefont {H.}~\bibnamefont
  {Monien}},\ }\href@noop {} {\bibfield  {journal} {\bibinfo  {journal}
  {\Jprb}\ }\textbf {\bibinfo {volume} {58}},\ \bibinfo {pages} {R14741}
  (\bibinfo {year} {1998})}\BibitemShut {NoStop}%
\bibitem [{\citenamefont {Ejima}\ \emph {et~al.}(2012)\citenamefont {Ejima},
  \citenamefont {Fehske}, \citenamefont {Gebhard}, \citenamefont
  {zu~M{\"u}nster}, \citenamefont {Knap}, \citenamefont {Arrigoni},\ and\
  \citenamefont {von~der Linden}}]{ejima2012characterization}%
  \BibitemOpen
  \bibfield  {author} {\bibinfo {author} {\bibfnamefont {S.}~\bibnamefont
  {Ejima}}, \bibinfo {author} {\bibfnamefont {H.}~\bibnamefont {Fehske}},
  \bibinfo {author} {\bibfnamefont {F.}~\bibnamefont {Gebhard}}, \bibinfo
  {author} {\bibfnamefont {K.}~\bibnamefont {zu~M{\"u}nster}}, \bibinfo
  {author} {\bibfnamefont {M.}~\bibnamefont {Knap}}, \bibinfo {author}
  {\bibfnamefont {E.}~\bibnamefont {Arrigoni}}, \ and\ \bibinfo {author}
  {\bibfnamefont {W.}~\bibnamefont {von~der Linden}},\ }\href@noop {}
  {\bibfield  {journal} {\bibinfo  {journal} {\Jpra}\ }\textbf {\bibinfo
  {volume} {85}},\ \bibinfo {pages} {053644} (\bibinfo {year}
  {2012})}\BibitemShut {NoStop}%
\bibitem [{\citenamefont {et~al.}(2018)}]{wang2018high}%
  \BibitemOpen
  \bibfield  {author} {\bibinfo {author} {\bibfnamefont {T.~Wang}\ \bibnamefont
  {et~al.}},\ }\href@noop {} {\bibfield  {journal} {\bibinfo  {journal}
  {\Jprb}\ }\textbf {\bibinfo {volume} {98}},\ \bibinfo {pages} {245107}
  (\bibinfo {year} {2018})}\BibitemShut {NoStop}%
\bibitem [{\citenamefont {Ceperley}(1995)}]{ceperley1995}%
  \BibitemOpen
  \bibfield  {author} {\bibinfo {author} {\bibfnamefont {D.~M.}\ \bibnamefont
  {Ceperley}},\ }\href@noop {} {\bibfield  {journal} {\bibinfo  {journal}
  {\Jrmp}\ }\textbf {\bibinfo {volume} {67}},\ \bibinfo {pages} {279--355}
  (\bibinfo {year} {1995})}\BibitemShut {NoStop}%
\bibitem [{\citenamefont {Boninsegni}\ \emph
  {et~al.}(2006{\natexlab{a}})\citenamefont {Boninsegni}, \citenamefont
  {Prokof'ev},\ and\ \citenamefont {Svistunov}}]{boninsegni2006a}%
  \BibitemOpen
  \bibfield  {author} {\bibinfo {author} {\bibfnamefont {M.}~\bibnamefont
  {Boninsegni}}, \bibinfo {author} {\bibfnamefont {N.}~\bibnamefont
  {Prokof'ev}}, \ and\ \bibinfo {author} {\bibfnamefont {B.}~\bibnamefont
  {Svistunov}},\ }\href@noop {} {\bibfield  {journal} {\bibinfo  {journal}
  {\Jprl}\ }\textbf {\bibinfo {volume} {96}},\ \bibinfo {pages} {070601}
  (\bibinfo {year} {2006}{\natexlab{a}})}\BibitemShut {NoStop}%
\bibitem [{\citenamefont {Boninsegni}\ \emph
  {et~al.}(2006{\natexlab{b}})\citenamefont {Boninsegni}, \citenamefont
  {Prokof'ev},\ and\ \citenamefont {Svistunov}}]{boninsegni2006b}%
  \BibitemOpen
  \bibfield  {author} {\bibinfo {author} {\bibfnamefont {M.}~\bibnamefont
  {Boninsegni}}, \bibinfo {author} {\bibfnamefont {N.~V.}\ \bibnamefont
  {Prokof'ev}}, \ and\ \bibinfo {author} {\bibfnamefont {B.~V.}\ \bibnamefont
  {Svistunov}},\ }\href@noop {} {\bibfield  {journal} {\bibinfo  {journal}
  {\Jpre}\ }\textbf {\bibinfo {volume} {74}},\ \bibinfo {pages} {036701}
  (\bibinfo {year} {2006}{\natexlab{b}})}\BibitemShut {NoStop}%
\bibitem [{\citenamefont {Yao}\ \emph {et~al.}(2018)\citenamefont {Yao},
  \citenamefont {Cl\'ement}, \citenamefont {Minguzzi}, \citenamefont
  {Vignolo},\ and\ \citenamefont {Sanchez-Palencia}}]{yao2018}%
  \BibitemOpen
  \bibfield  {author} {\bibinfo {author} {\bibfnamefont {H.}~\bibnamefont
  {Yao}}, \bibinfo {author} {\bibfnamefont {D.}~\bibnamefont {Cl\'ement}},
  \bibinfo {author} {\bibfnamefont {A.}~\bibnamefont {Minguzzi}}, \bibinfo
  {author} {\bibfnamefont {P.}~\bibnamefont {Vignolo}}, \ and\ \bibinfo
  {author} {\bibfnamefont {L.}~\bibnamefont {Sanchez-Palencia}},\ }\href@noop
  {} {\bibfield  {journal} {\bibinfo  {journal} {\Jprl}\ }\textbf {\bibinfo
  {volume} {121}},\ \bibinfo {pages} {220402} (\bibinfo {year}
  {2018})}\BibitemShut {NoStop}%
\bibitem [{\citenamefont {Yao}\ \emph {et~al.}(2020)\citenamefont {Yao},
  \citenamefont {Giamarchi},\ and\ \citenamefont {Sanchez-Palencia}}]{yao2020}%
  \BibitemOpen
  \bibfield  {author} {\bibinfo {author} {\bibfnamefont {H.}~\bibnamefont
  {Yao}}, \bibinfo {author} {\bibfnamefont {T.}~\bibnamefont {Giamarchi}}, \
  and\ \bibinfo {author} {\bibfnamefont {L.}~\bibnamefont {Sanchez-Palencia}},\
  }\href@noop {} {\bibfield  {journal} {\bibinfo  {journal} {\Jprl}\ }\textbf
  {\bibinfo {volume} {125}},\ \bibinfo {pages} {060401} (\bibinfo {year}
  {2020})}\BibitemShut {NoStop}%
\bibitem [{\citenamefont {Gautier}\ \emph {et~al.}(2021)\citenamefont
  {Gautier}, \citenamefont {Yao},\ and\ \citenamefont
  {Sanchez-Palencia}}]{gautier2021}%
  \BibitemOpen
  \bibfield  {author} {\bibinfo {author} {\bibfnamefont {R.}~\bibnamefont
  {Gautier}}, \bibinfo {author} {\bibfnamefont {H.}~\bibnamefont {Yao}}, \ and\
  \bibinfo {author} {\bibfnamefont {L.}~\bibnamefont {Sanchez-Palencia}},\
  }\href@noop {} {\bibfield  {journal} {\bibinfo  {journal} {\Jprl}\ }\textbf
  {\bibinfo {volume} {126}},\ \bibinfo {pages} {110401} (\bibinfo {year}
  {2021})}\BibitemShut {NoStop}%
\bibitem [{\citenamefont {L{\"u}hmann}\ \emph {et~al.}(2012)\citenamefont
  {L{\"u}hmann}, \citenamefont {J{\"u}rgensen},\ and\ \citenamefont
  {Sengstock}}]{luhmann2012multi}%
  \BibitemOpen
  \bibfield  {author} {\bibinfo {author} {\bibfnamefont {D.-S.}\ \bibnamefont
  {L{\"u}hmann}}, \bibinfo {author} {\bibfnamefont {O.}~\bibnamefont
  {J{\"u}rgensen}}, \ and\ \bibinfo {author} {\bibfnamefont {K.}~\bibnamefont
  {Sengstock}},\ }\href@noop {} {\bibfield  {journal} {\bibinfo  {journal} {New
  Journal of Physics}\ }\textbf {\bibinfo {volume} {14}},\ \bibinfo {pages}
  {033021} (\bibinfo {year} {2012})}\BibitemShut {NoStop}%
\bibitem [{\citenamefont {Dutta}\ \emph {et~al.}(2011)\citenamefont {Dutta},
  \citenamefont {Eckardt}, \citenamefont {Hauke}, \citenamefont {Malomed},\
  and\ \citenamefont {Lewenstein}}]{dutta2011bose}%
  \BibitemOpen
  \bibfield  {author} {\bibinfo {author} {\bibfnamefont {O.}~\bibnamefont
  {Dutta}}, \bibinfo {author} {\bibfnamefont {A.}~\bibnamefont {Eckardt}},
  \bibinfo {author} {\bibfnamefont {P.}~\bibnamefont {Hauke}}, \bibinfo
  {author} {\bibfnamefont {B.}~\bibnamefont {Malomed}}, \ and\ \bibinfo
  {author} {\bibfnamefont {M.}~\bibnamefont {Lewenstein}},\ }\href@noop {}
  {\bibfield  {journal} {\bibinfo  {journal} {New Journal of Physics}\ }\textbf
  {\bibinfo {volume} {13}},\ \bibinfo {pages} {023019} (\bibinfo {year}
  {2011})}\BibitemShut {NoStop}%
\bibitem [{\citenamefont {Bissbort}\ \emph {et~al.}(2012)\citenamefont
  {Bissbort}, \citenamefont {Deuretzbacher},\ and\ \citenamefont
  {Hofstetter}}]{bissbort2012effective}%
  \BibitemOpen
  \bibfield  {author} {\bibinfo {author} {\bibfnamefont {U.}~\bibnamefont
  {Bissbort}}, \bibinfo {author} {\bibfnamefont {F.}~\bibnamefont
  {Deuretzbacher}}, \ and\ \bibinfo {author} {\bibfnamefont {W.}~\bibnamefont
  {Hofstetter}},\ }\href@noop {} {\bibfield  {journal} {\bibinfo  {journal}
  {\Jpra}\ }\textbf {\bibinfo {volume} {86}},\ \bibinfo {pages} {023617}
  (\bibinfo {year} {2012})}\BibitemShut {NoStop}%
\bibitem [{\citenamefont {Wright}\ \emph {et~al.}(2008)\citenamefont {Wright},
  \citenamefont {Leslie},\ and\ \citenamefont {Bigelow}}]{wright2008raman}%
  \BibitemOpen
  \bibfield  {author} {\bibinfo {author} {\bibfnamefont {K.~C.}\ \bibnamefont
  {Wright}}, \bibinfo {author} {\bibfnamefont {L.~S.}\ \bibnamefont {Leslie}},
  \ and\ \bibinfo {author} {\bibfnamefont {N.~P.}\ \bibnamefont {Bigelow}},\
  }\href@noop {} {\bibfield  {journal} {\bibinfo  {journal} {\Jpra}\ }\textbf
  {\bibinfo {volume} {78}},\ \bibinfo {pages} {053412} (\bibinfo {year}
  {2008})}\BibitemShut {NoStop}%
\bibitem [{\citenamefont {Cohen-Tannoudji}\ \emph {et~al.}(1998)\citenamefont
  {Cohen-Tannoudji}, \citenamefont {Dupont-Roc},\ and\ \citenamefont
  {Grynberg}}]{cohen1998atom}%
  \BibitemOpen
  \bibfield  {author} {\bibinfo {author} {\bibfnamefont {C.}~\bibnamefont
  {Cohen-Tannoudji}}, \bibinfo {author} {\bibfnamefont {J.}~\bibnamefont
  {Dupont-Roc}}, \ and\ \bibinfo {author} {\bibfnamefont {G.}~\bibnamefont
  {Grynberg}},\ }\href@noop {} {\emph {\bibinfo {title} {Atom-photon
  interactions: basic processes and applications}}}\ (\bibinfo {year}
  {1998})\BibitemShut {NoStop}%
\bibitem [{\citenamefont {Torrontegui}\ \emph {et~al.}(2011)\citenamefont
  {Torrontegui}, \citenamefont {Ib{\'a}{\~n}ez}, \citenamefont {Chen},
  \citenamefont {Ruschhaupt}, \citenamefont {Gu{\'e}ry-Odelin},\ and\
  \citenamefont {Muga}}]{torrontegui2011fast}%
  \BibitemOpen
  \bibfield  {author} {\bibinfo {author} {\bibfnamefont {E.}~\bibnamefont
  {Torrontegui}}, \bibinfo {author} {\bibfnamefont {S.}~\bibnamefont
  {Ib{\'a}{\~n}ez}}, \bibinfo {author} {\bibfnamefont {X.}~\bibnamefont
  {Chen}}, \bibinfo {author} {\bibfnamefont {A.}~\bibnamefont {Ruschhaupt}},
  \bibinfo {author} {\bibfnamefont {D.}~\bibnamefont {Gu{\'e}ry-Odelin}}, \
  and\ \bibinfo {author} {\bibfnamefont {J.~G.}\ \bibnamefont {Muga}},\
  }\href@noop {} {\bibfield  {journal} {\bibinfo  {journal} {\Jpra}\ }\textbf
  {\bibinfo {volume} {83}},\ \bibinfo {pages} {013415} (\bibinfo {year}
  {2011})}\BibitemShut {NoStop}%
\bibitem [{\citenamefont {Corgier}\ \emph {et~al.}(2018)\citenamefont
  {Corgier}, \citenamefont {Amri}, \citenamefont {Herr}, \citenamefont
  {Ahlers}, \citenamefont {Rudolph}, \citenamefont {Gu{\'e}ry-Odelin},
  \citenamefont {Rasel}, \citenamefont {Charron},\ and\ \citenamefont
  {Gaaloul}}]{corgier2018fast}%
  \BibitemOpen
  \bibfield  {author} {\bibinfo {author} {\bibfnamefont {R.}~\bibnamefont
  {Corgier}}, \bibinfo {author} {\bibfnamefont {S.}~\bibnamefont {Amri}},
  \bibinfo {author} {\bibfnamefont {W.}~\bibnamefont {Herr}}, \bibinfo {author}
  {\bibfnamefont {H.}~\bibnamefont {Ahlers}}, \bibinfo {author} {\bibfnamefont
  {J.}~\bibnamefont {Rudolph}}, \bibinfo {author} {\bibfnamefont
  {D.}~\bibnamefont {Gu{\'e}ry-Odelin}}, \bibinfo {author} {\bibfnamefont
  {E.~M.}\ \bibnamefont {Rasel}}, \bibinfo {author} {\bibfnamefont
  {E.}~\bibnamefont {Charron}}, \ and\ \bibinfo {author} {\bibfnamefont
  {N.}~\bibnamefont {Gaaloul}},\ }\href@noop {} {\bibfield  {journal} {\bibinfo
   {journal} {New Journal of Physics}\ }\textbf {\bibinfo {volume} {20}},\
  \bibinfo {pages} {055002} (\bibinfo {year} {2018})}\BibitemShut {NoStop}%
\bibitem [{\citenamefont {Gu{\'e}ry-Odelin}\ \emph {et~al.}(2019)\citenamefont
  {Gu{\'e}ry-Odelin}, \citenamefont {Ruschhaupt}, \citenamefont {Kiely},
  \citenamefont {Torrontegui}, \citenamefont {Mart{\'\i}nez-Garaot},\ and\
  \citenamefont {Muga}}]{guery2019shortcuts}%
  \BibitemOpen
  \bibfield  {author} {\bibinfo {author} {\bibfnamefont {D.}~\bibnamefont
  {Gu{\'e}ry-Odelin}}, \bibinfo {author} {\bibfnamefont {A.}~\bibnamefont
  {Ruschhaupt}}, \bibinfo {author} {\bibfnamefont {A.}~\bibnamefont {Kiely}},
  \bibinfo {author} {\bibfnamefont {E.}~\bibnamefont {Torrontegui}}, \bibinfo
  {author} {\bibfnamefont {S.}~\bibnamefont {Mart{\'\i}nez-Garaot}}, \ and\
  \bibinfo {author} {\bibfnamefont {J.~G.}\ \bibnamefont {Muga}},\ }\href@noop
  {} {\bibfield  {journal} {\bibinfo  {journal} {Rev. Mod. Phys.}\ }\textbf
  {\bibinfo {volume} {91}},\ \bibinfo {pages} {045001} (\bibinfo {year}
  {2019})}\BibitemShut {NoStop}%
\bibitem [{\citenamefont {Ding}\ \emph {et~al.}(2020)\citenamefont {Ding},
  \citenamefont {Huang}, \citenamefont {Paul}, \citenamefont {Hao},\ and\
  \citenamefont {Chen}}]{ding2020smooth}%
  \BibitemOpen
  \bibfield  {author} {\bibinfo {author} {\bibfnamefont {Y.}~\bibnamefont
  {Ding}}, \bibinfo {author} {\bibfnamefont {T.-Y.}\ \bibnamefont {Huang}},
  \bibinfo {author} {\bibfnamefont {K.}~\bibnamefont {Paul}}, \bibinfo {author}
  {\bibfnamefont {M.}~\bibnamefont {Hao}}, \ and\ \bibinfo {author}
  {\bibfnamefont {X.}~\bibnamefont {Chen}},\ }\href@noop {} {\bibfield
  {journal} {\bibinfo  {journal} {\Jpra}\ }\textbf {\bibinfo {volume} {101}},\
  \bibinfo {pages} {063410} (\bibinfo {year} {2020})}\BibitemShut {NoStop}%
\bibitem [{\citenamefont {Gerbier}\ \emph {et~al.}(2005)\citenamefont
  {Gerbier}, \citenamefont {Widera}, \citenamefont {F{\"o}lling}, \citenamefont
  {Mandel}, \citenamefont {Gericke},\ and\ \citenamefont
  {Bloch}}]{gerbier2005interference}%
  \BibitemOpen
  \bibfield  {author} {\bibinfo {author} {\bibfnamefont {F.}~\bibnamefont
  {Gerbier}}, \bibinfo {author} {\bibfnamefont {A.}~\bibnamefont {Widera}},
  \bibinfo {author} {\bibfnamefont {S.}~\bibnamefont {F{\"o}lling}}, \bibinfo
  {author} {\bibfnamefont {O.}~\bibnamefont {Mandel}}, \bibinfo {author}
  {\bibfnamefont {T.}~\bibnamefont {Gericke}}, \ and\ \bibinfo {author}
  {\bibfnamefont {I.}~\bibnamefont {Bloch}},\ }\href@noop {} {\bibfield
  {journal} {\bibinfo  {journal} {\Jpra}\ }\textbf {\bibinfo {volume} {72}},\
  \bibinfo {pages} {053606} (\bibinfo {year} {2005})}\BibitemShut {NoStop}%
\bibitem [{\citenamefont {Celi}\ \emph {et~al.}(2014)\citenamefont {Celi},
  \citenamefont {Massignan}, \citenamefont {Ruseckas}, \citenamefont {Goldman},
  \citenamefont {Spielman}, \citenamefont {Juzeli{\=u}nas},\ and\ \citenamefont
  {Lewenstein}}]{celi2014synthetic}%
  \BibitemOpen
  \bibfield  {author} {\bibinfo {author} {\bibfnamefont {A.}~\bibnamefont
  {Celi}}, \bibinfo {author} {\bibfnamefont {P.}~\bibnamefont {Massignan}},
  \bibinfo {author} {\bibfnamefont {J.}~\bibnamefont {Ruseckas}}, \bibinfo
  {author} {\bibfnamefont {N.}~\bibnamefont {Goldman}}, \bibinfo {author}
  {\bibfnamefont {I.~B.}\ \bibnamefont {Spielman}}, \bibinfo {author}
  {\bibfnamefont {G.}~\bibnamefont {Juzeli{\=u}nas}}, \ and\ \bibinfo {author}
  {\bibfnamefont {M.}~\bibnamefont {Lewenstein}},\ }\href@noop {} {\bibfield
  {journal} {\bibinfo  {journal} {\Jprl}\ }\textbf {\bibinfo {volume} {112}},\
  \bibinfo {pages} {043001} (\bibinfo {year} {2014})}\BibitemShut {NoStop}%
\bibitem [{\citenamefont {Taddia}\ \emph {et~al.}(2017)\citenamefont {Taddia},
  \citenamefont {Cornfeld}, \citenamefont {Rossini}, \citenamefont {Mazza},
  \citenamefont {Sela},\ and\ \citenamefont {Fazio}}]{taddia2017topological}%
  \BibitemOpen
  \bibfield  {author} {\bibinfo {author} {\bibfnamefont {L.}~\bibnamefont
  {Taddia}}, \bibinfo {author} {\bibfnamefont {E.}~\bibnamefont {Cornfeld}},
  \bibinfo {author} {\bibfnamefont {D.}~\bibnamefont {Rossini}}, \bibinfo
  {author} {\bibfnamefont {L.}~\bibnamefont {Mazza}}, \bibinfo {author}
  {\bibfnamefont {E.}~\bibnamefont {Sela}}, \ and\ \bibinfo {author}
  {\bibfnamefont {R.}~\bibnamefont {Fazio}},\ }\href@noop {} {\bibfield
  {journal} {\bibinfo  {journal} {\Jprl}\ }\textbf {\bibinfo {volume} {118}},\
  \bibinfo {pages} {230402} (\bibinfo {year} {2017})}\BibitemShut {NoStop}%
\bibitem [{\citenamefont {Price}\ \emph {et~al.}(2017)\citenamefont {Price},
  \citenamefont {Ozawa},\ and\ \citenamefont {Goldman}}]{price2017synthetic}%
  \BibitemOpen
  \bibfield  {author} {\bibinfo {author} {\bibfnamefont {H.~M.}\ \bibnamefont
  {Price}}, \bibinfo {author} {\bibfnamefont {T.}~\bibnamefont {Ozawa}}, \ and\
  \bibinfo {author} {\bibfnamefont {N.}~\bibnamefont {Goldman}},\ }\href@noop
  {} {\bibfield  {journal} {\bibinfo  {journal} {\Jpra}\ }\textbf {\bibinfo
  {volume} {95}},\ \bibinfo {pages} {023607} (\bibinfo {year}
  {2017})}\BibitemShut {NoStop}%
\bibitem [{\citenamefont {Troyer}\ \emph {et~al.}(1998)\citenamefont {Troyer},
  \citenamefont {Ammon},\ and\ \citenamefont {Heeb}}]{troyer1998}%
  \BibitemOpen
  \bibfield  {author} {\bibinfo {author} {\bibfnamefont {M.}~\bibnamefont
  {Troyer}}, \bibinfo {author} {\bibfnamefont {B.}~\bibnamefont {Ammon}}, \
  and\ \bibinfo {author} {\bibfnamefont {E.}~\bibnamefont {Heeb}},\ }\href@noop
  {} {\bibfield  {journal} {\bibinfo  {journal} {Lect. Notes Comput. Sci.}\
  }\textbf {\bibinfo {volume} {1505}},\ \bibinfo {pages} {191} (\bibinfo {year}
  {1998})}\BibitemShut {NoStop}%
\bibitem [{\citenamefont {et~al.}(2007)}]{ALPS2007}%
  \BibitemOpen
  \bibfield  {author} {\bibinfo {author} {\bibfnamefont {A.~F.~Albuquerque}\
  \bibnamefont {et~al.}},\ }\href@noop {} {\bibfield  {journal} {\bibinfo
  {journal} {J. Magn. Magn. Mater.}\ }\textbf {\bibinfo {volume} {310}},\
  \bibinfo {pages} {1187} (\bibinfo {year} {2007})}\BibitemShut {NoStop}%
\bibitem [{\citenamefont {et~al.}(2011)}]{ALPS2011}%
  \BibitemOpen
  \bibfield  {author} {\bibinfo {author} {\bibfnamefont {B.~Bauer}\
  \bibnamefont {et~al.}},\ }\href@noop {} {\bibfield  {journal} {\bibinfo
  {journal} {J. Stat. Mech.: Th. Exp.}\ }\textbf {\bibinfo {volume} {05}},\
  \bibinfo {pages} {P05001} (\bibinfo {year} {2011})}\BibitemShut {NoStop}%
\bibitem [{\citenamefont {Grimm}\ \emph {et~al.}(2000)\citenamefont {Grimm},
  \citenamefont {Weidem{\"u}ller},\ and\ \citenamefont
  {Ovchinnikov}}]{grimm2000optical}%
  \BibitemOpen
  \bibfield  {author} {\bibinfo {author} {\bibfnamefont {R.}~\bibnamefont
  {Grimm}}, \bibinfo {author} {\bibfnamefont {M.}~\bibnamefont
  {Weidem{\"u}ller}}, \ and\ \bibinfo {author} {\bibfnamefont {Y.~B.}\
  \bibnamefont {Ovchinnikov}},\ }\href@noop {} {\bibfield  {journal} {\bibinfo
  {journal} {Advances in atomic, molecular, and optical physics}\ }\textbf
  {\bibinfo {volume} {42}},\ \bibinfo {pages} {95--170} (\bibinfo {year}
  {2000})}\BibitemShut {NoStop}%
\end{thebibliography}%



 \renewcommand{\theequation}{S\arabic{equation}}

 \setcounter{equation}{0}

 \renewcommand{\thefigure}{S\arabic{figure}}

 \setcounter{figure}{0}

 \renewcommand{\thesection}{S\arabic{section}}

 \setcounter{section}{0}

 \onecolumngrid

 %\pagebreak

 \newpage

{\center \bf \large Supplemental Material for \\}

{\center \bf \large A momentum dependent optical lattice induced by artificial gauge potential \\ \vspace*{1.cm}
}

In this supplemental material, we provide details about zeroth-order component of the Floquet Hamiltonian and the adiabatic condition calculations (Sec.~\ref{sec:Floquet}), the single particle ground state of the TMD optical lattice (Sec.~\ref{sec:SingleParticle}), the $\bar{n}=2$ Mott lobe calculation result (Sec.~\ref{sec:n=2}), and the detailed construction of the periodically driven Hamiltonian via a Raman process (Sec.~\ref{sec:Raman}).

\section{Zeroth-order component of the Floquet Hamiltonian and the adiabatic condition}
\label{sec:Floquet}

In this work, the system we consider satisfies the Floquet adiabatic condition in the theoretical formalism of the main text. We can rewrite eqn.(2) in the main text in the form
\begin{equation}
H_F=\frac{\vec{p}^2}{2M}-\frac{1}{2M}(\vec{A}\cdot\vec{p}+\vec{p}\cdot\vec{A})+\frac{\vec{A}^2}{2M}+V_{ext}(\vec{r}).
\end{equation}
The adiabatic condition can be rewritten as
\begin{equation}\label{adiabatic condition-2}
|\langle \vec{k}|[\frac{1}{2M}(\vec{A}\cdot\vec{p}+\vec{p}\cdot\vec{A})+\frac{\vec{A}^2}{2M}]^{(n)}|\vec{k}^{'}\rangle|\ll\hbar\omega (n\ne0),
\end{equation}
where $[\cdots]^{(n)}$ is the $n$-th Fourier component. From eqn.(\ref{adiabatic condition-2}), we can see that the adiabatic condition depends not only on the ratio $\Omega_0/\omega$, but also on the momentum of the particle. A rigorous solution of the adiabatic condition can be lengthy, so we focus on a operational condition.

We first work on the term $\vec{A}^2/2M$. Writing down the analytic form of the vector potential, $\vec{A}$ \cite{ravckauskas2019non,novivcenko2019non,chen20202}:
\begin{equation}\label{analytic vector potential full}
A_{\xi}=d_{1\xi}+d_{2\xi}+d_{3\xi},
\end{equation}
where
\begin{eqnarray}\label{analytic vector potential components}
&&d_{1\xi}=\hbar a\sin{\omega t}\frac{(\vec{\Omega}\cdot\partial_{\xi}\vec{\Omega})(\vec{\Omega}\cdot\vec{\sigma})}{2|\Omega|^3}, \nonumber \\
&&d_{2\xi}=\hbar \sin{(a\sin{\omega t})}\frac{[(\vec{\Omega}\times\partial_{\xi}\vec{\Omega})\times\vec{\Omega}]\cdot\vec{\sigma}}{2|\Omega|^3}, \\
&&d_{3\xi}=\hbar[\cos{(a\sin{\omega t})}-1]\frac{\vec{\Omega}\times\partial_{\xi}\vec{\Omega}\cdot\vec{\sigma}}{2|\Omega|^2}, \nonumber
\end{eqnarray}
are the $\xi$-th component of vectors $\vec{d}_1$, $\vec{d}_2$ and $\vec{d}_3$. Then,
\begin{equation}\label{A_xi2}
A_{\xi}^2=d_{1\xi}^2+d_{2\xi}^2+d_{3\xi}^2+(d_{1\xi}d_{2\xi}+d_{2\xi}d_{1\xi})+(d_{1\xi}d_{3\xi}+d_{3\xi}d_{1\xi})+(d_{2\xi}d_{3\xi}+d_{3\xi}d_{2\xi}).
\end{equation}
We can simplify eqn.(\ref{A_xi2}) by noticing that the last three terms in the parenthesis vanish:
\begin{eqnarray}
(d_{1\xi}d_{2\xi}+d_{2\xi}d_{1\xi})&\propto&\Omega_i\partial_{\xi}\Omega_i\Omega_j\hat{\sigma}_j\epsilon_{klm}C_{\xi k}\Omega_l\hat{\sigma}_m+\epsilon_{klm}C_{\xi k}\Omega_l\hat{\sigma}_m\Omega_i\partial_{\xi}\Omega_i\Omega_j\hat{\sigma}_j \nonumber \\
&=&\Omega_i\partial_{\xi}\Omega_i\Omega_j\epsilon_{klm}C_{\xi k}\Omega_l(\hat{\sigma}_j\hat{\sigma}_m+\hat{\sigma}_m\hat{\sigma}_j) \nonumber \\
&=&0, \nonumber \\
(d_{2\xi}d_{3\xi}+d_{3\xi}d_{2\xi})&\propto&\epsilon_{ijk}C_{\xi i}\Omega_j\hat{\sigma}_k C_{\xi l}\hat{\sigma}_l+C_{\xi l}\hat{\sigma}_l\epsilon_{ijk}C_{\xi i}\Omega_j\hat{\sigma}_k \nonumber \\
&=&\epsilon_{ijk}C_{\xi i}\Omega_j C_{\xi l}(\hat{\sigma}_k\hat{\sigma}_l+\hat{\sigma}_l\hat{\sigma}_k) \nonumber \\
&=&0, \\
(d_{1\xi}d_{3\xi}+d_{3\xi}d_{1\xi})&\propto&\Omega_i\partial_{\xi}\Omega_i\Omega_j\hat{\sigma}_j C_{\xi k}\hat{\sigma}_k+C_{\xi k}\hat{\sigma}_k\Omega_i\partial_{\xi}\Omega_i\Omega_j\hat{\sigma}_j \nonumber \\
&=&\Omega_i\partial_{\xi}\Omega_i\Omega_jC_{\xi k}(\hat{\sigma}_j\hat{\sigma}_k+\hat{\sigma}_k\hat{\sigma}_j) \nonumber \\
&=&2\Omega_i\partial_{\xi}\Omega_i\Omega_j\epsilon_{jlm}\Omega_l\partial_{\xi}\Omega_m \nonumber \\
&=&0, \nonumber
\end{eqnarray}
where we let $C_{\xi i}=\{\vec{\Omega}\times\partial_{\xi}\vec{\Omega}\}_{i}=\epsilon_{ijk}\Omega_j\partial_{\xi}\Omega_k$ and we used the property $\hat{\sigma}_i\hat{\sigma}_j+\hat{\sigma}_j\hat{\sigma}_i=2\delta_{ij}\mathbbm{1}$.

After the simplification above, eqn.(\ref{A_xi2}) becomes
\begin{equation}\label{A_mu2 analytical}
A_{\xi}^2=\hbar^2a^2\sin^{2}{\omega t}\frac{(\vec{\Omega}\cdot\partial_{\xi}\vec{\Omega})^2}{4|\Omega|^4}+2\hbar^2[1-\cos{(a\sin{\omega t)}}]\frac{|\Omega|^2(\partial_{\xi}\vec{\Omega})^2-(\vec{\Omega}\cdot\partial_{\xi}\vec{\Omega})^2}{4|\Omega|^4}.
\end{equation} 
We can write $\cos{(a\sin{\omega t})}$ as an expansion of Bessel functions of the first kind,
\begin{equation}
\cos{(a\sin{\omega t})}=J_0(a)+2\sum_{m=1}^{\infty}J_{2m}(a)\cos{(2m\omega t)}.
\end{equation}
The zeroth order Fourier term of $\vec{A}^2/2M$ is
\begin{equation}\label{A2 zeroth}
\left[\frac{\vec{A}^2}{2M}\right]^{(0)}=\{\frac{1}{16}a_0^2(1-\cos{2k_Lx})+\frac{1}{2}\left[1-J_0(a)\right]\}E_r,
\end{equation}
and the second order ($\pm2$) harmonic terms are
\begin{equation}\label{A2 2nd order}
\left[\frac{\vec{A}^2}{2M}\right]^{(\pm2)}=[\frac{1}{32}a_0^2\cos{2k_{L}x}\pm\frac{i}{2}J_2(a)]E_r.
\end{equation}

Next, we calculate the contribution of the $\frac{1}{2M}(\vec{A}\cdot\vec{p}+\vec{p}\cdot\vec{A})$ term. From  eqn.(\ref{analytic vector potential components}), we get
\begin{eqnarray}\label{exact gauge potetnial}
\vec{A}&=&\frac{1}{2}\hbar k_L\{[-a_0\sin{k_L x}e^{-ik_Lz}\sin{\omega t}\hat{x}+ie^{ik_Lz}\sin{(a\sin{\omega t})}sgn(a)\hat{z}]\hat{\sigma}_{+}+h.c.\} \nonumber \\
&+&\frac{1}{2}\hbar k_L\left[\cos{(a\sin{\omega t})}-1\right]\hat{\sigma}_{3}\hat{z}, 
\end{eqnarray}
where $sgn(\cdot)$ is the signum function, $\hat{\sigma}_{+}=(\hat{\sigma}_{1}-i\hat{\sigma}_{2})/\sqrt{2}$, $\hat{x}$ and $\hat{z}$ are unit vector along the $x$ and $z$-axes. Then, we can write down the absolute value of the non-zero matrix element of the Floquet Hamiltonian that is proportional to the first Fourier order,
\begin{eqnarray}
\bra{k_x,k_z,\uparrow}H_F^{(\pm1)}\ket{k_x^{'},k_z^{'},\downarrow}&=&\pm\frac{a_0}{8}E_r\{\bra{k_x,k_z,\uparrow}[(e^{ik_Lx}-e^{-ik_Lx})e^{-ik_Lz}\frac{k_x^{'}}{k_L}]\ket{k_x^{'},k_z^{'},\downarrow} \nonumber \\
&+&\bra{k_x,k_z,\uparrow}[\frac{k_x}{k_L}(e^{ik_Lx}-e^{-ik_Lx})e^{ik_Lz}]\ket{k_x^{'},k_z^{'},\downarrow}\} \nonumber \\
&\pm&\frac{E_r}{2}\bra{k_x,k_z,\uparrow}J_{1}(a)sgn(a)(e^{ ik_Lz}\frac{k_z}{k_L}\mp e^{- ik_Lz}\frac{k_z^{'}}{k_L})\ket{k_x^{'},k_z^{'},\downarrow}, \nonumber
\end{eqnarray}
where $\ket{\uparrow(\downarrow)}$ indicates two spin states. Notice that the above equation can be split into several matrix elements between different momentum states. After collecting terms and simplifying, we get the adiabatic condition for the first Fourier order as
\begin{equation}\label{first order adiabatic condition}
\{\frac{a_0}{8}|\frac{k_x}{k_L}|+\frac{1}{2}|\langle k_L|sgn(a)J_{1}(a)|k_L\pm1\rangle||\frac{k_z}{k_L}|\}E_r=\frac{1}{2}\{|\frac{k_x}{k_L}|+0.19|\frac{k_z}{k_L}|\}E_r
\ll\hbar\omega,
\end{equation}
where we evaluated $|\langle k_L|sgn(a)J_{1}(a)|k_L\pm1\rangle|$ numerically to be 0.19 by setting $a_0=4$. In our calculations, we use $\hbar\omega/E_r=20$, and we can get the adiabatic condition for the first Fourier order by combining the two conditions in eqn.(\ref{first order adiabatic condition}) to be $|k_x/k_L|+0.19|k_z/k_L|\ll40$. We also evaluated the matrix elements between higher order momentum states and found them to be much smaller than the matrix element in eqn.(\ref{first order adiabatic condition}).

Next, we consider the second order Fourier components. Both the $\frac{1}{2M}(\vec{A}\cdot\vec{p}+\vec{p}\cdot\vec{A})$ and $A^2$ terms contribute to the even order $\sin{2m\omega t}$ terms. Specifically, for the second order, from eqn.(\ref{A2 2nd order}) and eqn.(\ref{exact gauge potetnial}) we can get the matrix element of second Fourier component,
\begin{eqnarray}
\langle k_x,k_z,\uparrow(\downarrow)|H_F^{(\pm2)}|k_x^{'},k_z,\uparrow(\downarrow)\rangle&=&\langle k_x,k_z,\uparrow(\downarrow)|\left[\frac{1}{64}a_0^2\pm\frac{i}{2}J_{2}(a)\right]E_r|k_x^{'}\pm2k_L,k_z,\uparrow(\downarrow)\rangle \nonumber \\
&+&\langle k_x,k_z,\uparrow(\downarrow)|2\frac{k_z}{k_L}J_{2}(a)E_r|k_x^{'}\pm2k_L,k_z,\uparrow(\downarrow)\rangle \\
&\approx&(\frac{1}{64}a_0^2\pm0.1i-0.4\frac{k_z}{k_L})E_r\delta(k_x-k_x^{'}\pm2k_L). \nonumber
\end{eqnarray}
where we used $\langle k_x|J_2(a)|k_x\pm 2k_L\rangle=-0.20$ from numerical evaluation using the same parameters as before.The adiabatic condition for the second Fourier order becomes $|(\frac{1}{64}a_0^2\pm0.1i-0.4\frac{k_z}{k_L})E_r|\ll\hbar\omega$, and we get $|k_z/k_L|\ll49$. Combine this result with $|k_x/k_L|+0.19|k_z/k_L|\ll40$, we get $|k_x/k_L|\ll31$. 

For higher order Fourier terms, one can do the calculation and see that the adiabatic conditions are all weaker than the first two Fourier terms. In our calculations, the above adiabatic conditions, $|k_x/k_L|\ll31$ and $|k_z/k_L|\ll49$, can easily be satisfied.

\section{Single particle ground state of the TMD optical lattice}
\label{sec:SingleParticle}
In the main text, the single particle Hamiltonian takes the form
\begin{equation}
H_{L}=\frac{\vec{p}^2}{2M}+V(x,p_z)+V_{ext}(\vec{r}), \nonumber
\end{equation}
where $V(x,p_z)=\{(p_z+1/2)\left[1-J_0(a)\right]+a_0^2(1-\cos{2k_Lx})/16\}E_r$. We can expand Hamiltonian $H_L$ in the momentum basis, $\ket{\pm2l\hbar k_L}$ ($l=0,\pm1,\pm2,\cdots$), and diagonalize the Hamiltonian to get the single-particle band structure and eigenstates. In all our calculations above, we have used the expansion form of the Bessel function of the first kind, $J_{q}(a)=\sum_{m=0}^{\infty}(-1)^m(a)^{2m+q}/2^{2m+q}\Gamma(m+1)\Gamma(m+q+1)$, where $q$ indicates the order of Bessel function. In practice, we truncate the expansion at the order of $l=\pm10$, which is accurate enough with our choice of parameters.

\begin{figure}
	\begin{center}
		\includegraphics[scale=0.35]{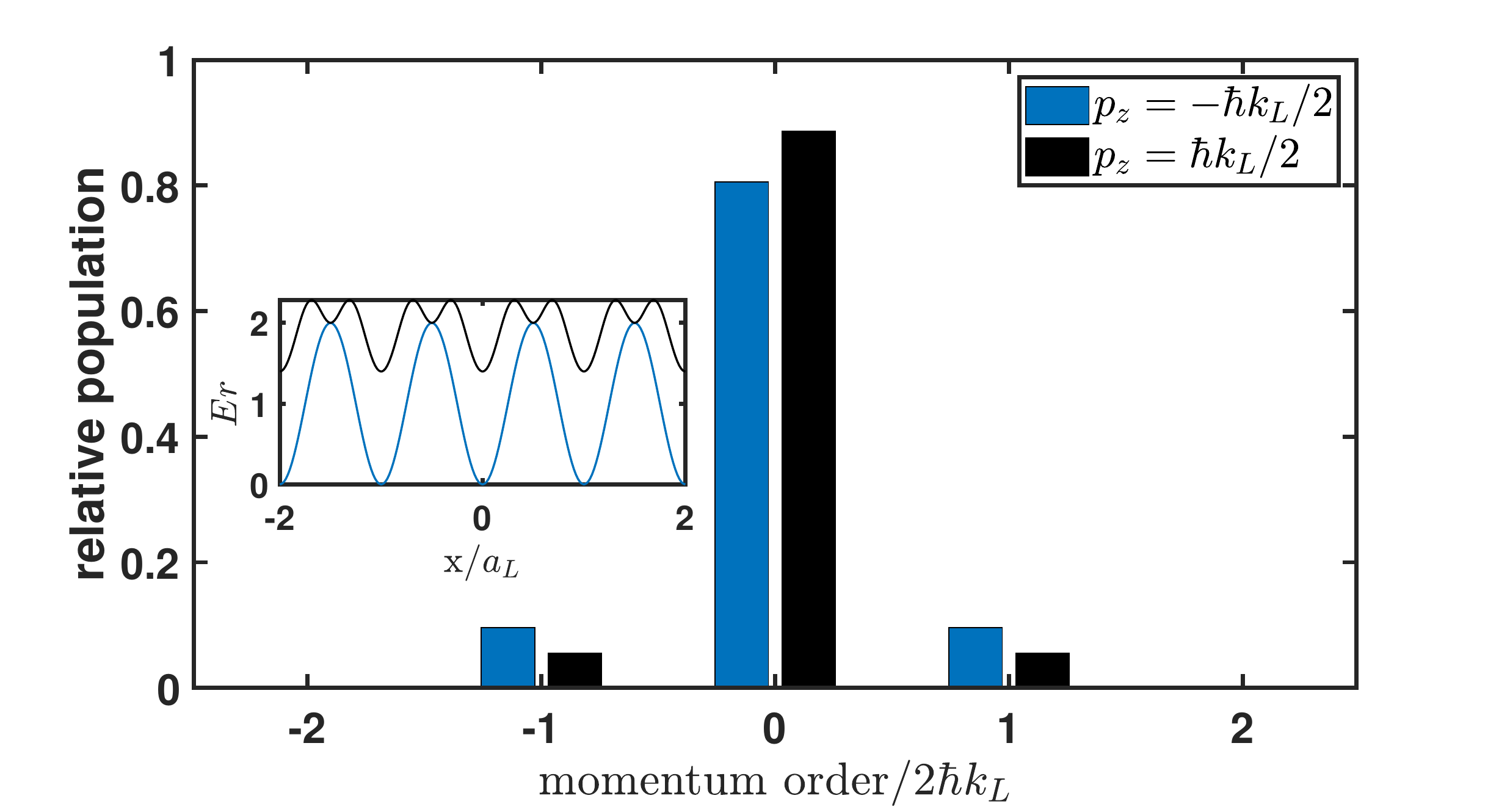}
	\end{center}
	\caption{$s$-band relative population of each momentum state along the $x$-axis with different $p_z$ at zero quasi-momentum. The inset plot is the corresponding $p_z$-dependent 1D optical lattice potential.}\label{single particle example}
\end{figure}

Consider the simplest case where $V_{ext}(\vec{r})=0$. When $p_z$ is low the on-site interaction in such a 1D optical lattice will be negligible. In such a weak interaction regime, we can study the physics of the system by solving the single-particle Hamiltonian, $H_L$. As shown in Fig.\ref{single particle example}, the 1D lattice potential is determined by $p_z$, which can be experimentally observed by measuring the population of each momentum state along the $x$-direction.

\section{Phase diagram around $\bar{n}=2$ Mott lobe}
\label{sec:n=2}

\begin{figure}
	\begin{center}
		\includegraphics[scale=0.3]{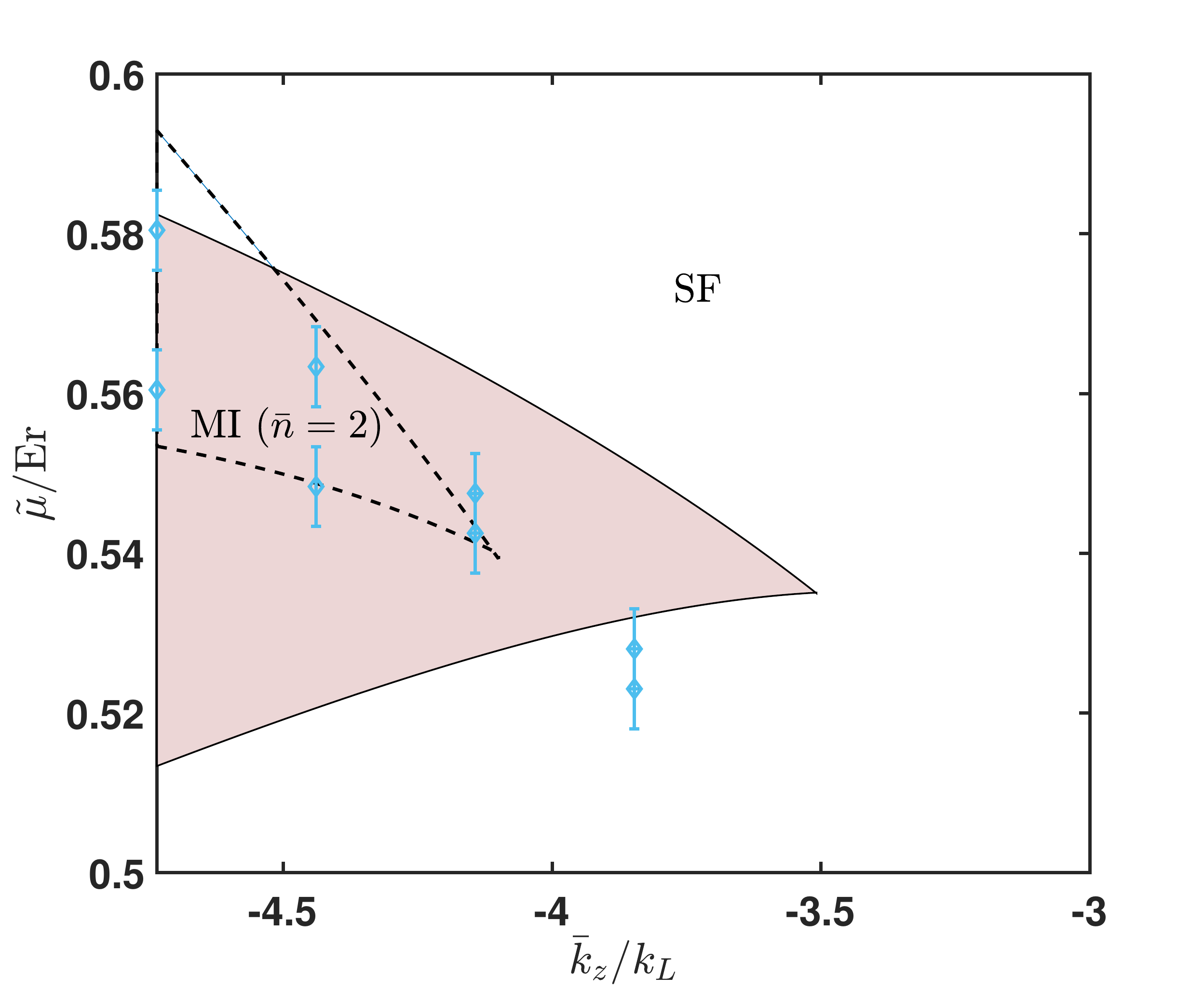}
	\end{center}
	\caption{$\bar{n}=2$ Mott lobe. The pink shaded area with black solid line boundaries is calculated from the occupation-independent Bose-Hubbard model, blue diamonds with errorbars ($\pm0.005Er$) are the QMC calculation results, and black dashed lines are Bose-Hubbard model calculation results with a band mixing ground state $|\psi_{g}\rangle\approx0.994|s\rangle+0.109|d\rangle$.}\label{n=2 Mott}
\end{figure}

In Fig.2 of the main text, we showed the phase diagram of our effective 1D optical lattice and found good agreement on the $\bar{n}=1$ Mott lobe between our effective 1D Bose-Hubbard model and the QMC calculation. In Fig.\ref{n=2 Mott} we show the $\bar{n}=2$ Mott lobe. There is a significant difference between the Bose-Hubbard model (pink shaded area with black solid line boundaries) and the QMC calculations (blue diamonds with errorbars). We believe that the difference is caused by the occupation-dependence of the tunneling and interactions terms, which makes the occupation-independent Bose-Hubbard model less accurate. Our QMC calculation is more accurate because we used a continuous potential in the QMC calculations instead of the Bose-Hubbard model. In previous work, occupation-dependent tunneling and onsite interactions in a modified Bose-Hubbard model have been studied \cite{luhmann2012multi,dutta2011bose,bissbort2012effective}, and the true many-body ground state was found to be a band mixing state. Phenomenologically, in our calculations we find a mixing state with $|\psi_{g}\rangle\approx0.994|s\rangle+0.109|d\rangle$ gives a better agreement between a Bose-Hubbard model with modified parameters and the QMC results (black dashed lines), where $|s\rangle$ and $|d\rangle$ are the single-particle eigenstates of the $s$ and $d$-bands, respectively. Our result agrees with the finding in Ref.\cite{luhmann2012multi} that the population on higher orbitals is smaller than $1\%$. Although for more rigorous analysis one would need to numerically diagonalize the many-body Hamiltonian \cite{luhmann2012multi,dutta2011bose,bissbort2012effective}, we believe that our qualitative calculations are good enough to understand the disagreement on the $\bar{n}=2$ Mott lobe.

\section{Detailed construction of the periodically driven Hamiltonian via Raman process and lifetime estimation}
\label{sec:Raman}

Using second order perturbation theory, we can calculate the effective Hamiltonian of the ground state manifold of the atom from the multi-laser Raman process Hamiltonian \cite{cohen1998atom,wright2008raman}. Here, we consider $^{87}$Rb atoms in the $|5S_{\frac{1}{2}},F=1\rangle$ manifold to be our system. More specifically, we consider $|1\rangle=|F=1,m_F=-1\rangle$, $|2\rangle=|F=1,m_F=0\rangle$ and $|3\rangle=|F=1,m_F=1\rangle$. By applying a bias magnetic field, $\vec{B}$, we can define the $+z$-direction. We then choose our Raman laser $a$ to be a standing wave along $x$-axis with $\pi$-polarization, laser $b$ propagates along $+z$-direction with $\sigma^{-}$-polarization, and laser $c$ is chosen to be a standing wave along the $x$-axis with $\pi$-polarization that is oppositely detuned compared to laser $a$ to cancel the spatially dependent AC stark shift. Specifically, lasers $a$ and $b$ are blue detuned from the $^{87}$Rb $D_1$ transition and laser $c$ is red detuned from $D_1$ transition. All the wavelengths of the Raman lasers considered here can be achieved in the laboratory (e.g.with a Ti:Sapphire laser). All one-photon detunings are much larger than Rabi frequencies so that there is negligible transitions to excited states. The level diagram is shown in Fig.3 in the main text. We only consider $D_1$ and $D_2$ lines and all other atomic states are too far from the laser frequency we consider so that the laser couplings to those states are negligible.

Similar to the calculations in previous works\cite{wright2008raman,chen20202}, we can write the reduced 3-level effective Hamiltonian for the Raman process with the laser fields in eqn.(9) of the main text,
\begin{eqnarray}\label{Raman Hamiltonian}
H_R&&=\sum_{\mathcal{D},F'}\sum_{m_F=-1}^{1}\{[ \frac{(c_{\mathcal{D},F,F'}^{m_F,m_F})^2}{\Delta_{a,F,F',\mathcal{D}}^{m_F,m_F}}|\Omega_{a,\mathcal{D}}|^2 +\frac{(c_{\mathcal{D},F,F'}^{m_F-1,m_F})^2}{\Delta_{b,F,F',\mathcal{D}}^{m_F,m_F}}|\Omega_{b,\mathcal{D}}|^2+\frac{(c_{\mathcal{D},F,F'}^{m_F,m_F})^2}{\Delta_{c,F,F',\mathcal{D}}^{m_F,m_F}}|\Omega_{c,\mathcal{D}}|^2 + \delta_{m_F} ] |m_F\rangle\langle m_F| \nonumber \\
&&+\frac{1}{2}(\frac{1}{\Delta_{a,F,F',\mathcal{D}}^{m_F,m_F}}+\frac{1}{\Delta_{b,F,F',\mathcal{D}}^{m_F+1,m_F}})c_{\mathcal{D},F,F'}^{m_F,m_F}c_{\mathcal{D},F,F'}^{m_F-1,m_F}\Omega_{a,\mathcal{D}}^{\ast}\Omega_{b,\mathcal{D}}|m_F\rangle\langle m_F+1| + h.c.  \},
\end{eqnarray}
where $F=1$, $\mathcal{D}=1,2$ indicates the transitions for $D_1$ and $D_2$ lines, $F'=1,2$ and $F'=0,1,2$ for $\mathcal{D}=1$ and $\mathcal{D}=2$, respectively. $c_{\mathcal{D},F,F'}^{m_F,m_{F'}}$ is the Clebsh-Gordon coefficient between $|F,m_F\rangle$ and $|F',m_{F'}\rangle$ states for $D_1$ ($D_2$) line and $\Delta_{\zeta,F,F',\mathcal{D}}^{m_F,m_{F'}}=\omega_{\zeta}-(\omega_{F',m_{F'}}^{e}-\omega_{F,m_F}^{g})$ ($\zeta=a,b,c$) is the one-photon detuning, where $\omega_{\zeta}$, $\omega_{F,m_F}^{g}$ and $\omega_{F',m_{F'}^{e}}$ are angular frequencies for laser $\zeta=a,b,c$ and the corresponding energy levels of ground state and excited state, respectively. $\Omega_{\zeta,\mathcal{D}}=-\vec{d}_{\mathcal{D}}\cdot\vec{E}_{\zeta}/\hbar$ ($\zeta=a,b,c$) are the Rabi frequencies, where $\vec{d}_{\mathcal{D}}$ ($\mathcal{D}=1,2$) are the effective dipole moment vector of $D_1$ and $D_2$ transition, respectively. $|m_F\rangle=|F=1,m_F=-1,0,1\rangle$ correspond to $|1\rangle$, $|2\rangle$ and $|3\rangle$, respectively. $\delta_{m_F}$ ($m_F=-1,0,1$) are defined as $\delta_{-1}=0$, $\delta_{0}=\Delta_{a,F,F',\mathcal{D}}^{-1,-1}-\Delta_{b,F,F',\mathcal{D}}^{-1,0}$ and $\delta_{1}=\Delta_{a,F,F',\mathcal{D}}^{-1,-1}-\Delta_{b,F,F',\mathcal{D}}^{-1,0}+\Delta_{a,F,F',\mathcal{D}}^{0,0}-\Delta_{b,F,F',\mathcal{D}}^{0,1}$. Since laser $a$ and $b$ are blue detuned from $D_1$ line while laser $c$ is red detuned from $D_1$ line, laser $c$ only contributes to the AC stark shift terms in diagonal matrix elements. If we rewrite eqn.(\ref{Raman Hamiltonian}) as
\begin{equation}
H_{R}=H_{11}|1\rangle\langle 1|+H_{22}|2\rangle\langle 2|+H_{33}|3\rangle\langle 3|+H_{12}|1\rangle\langle 2|+H_{23}|2\rangle\langle 3|+h.c., \nonumber
\end{equation}
and if $H_{22}=H_{11}$ and $|H_{33}-H_{22}|\gg H_{23}$, then $|1\rangle$ and $|2\rangle$ are well isolated from the third state, $|3\rangle$, and the effective two-level Hamiltonian takes the form $H_e=H_{11}|1\rangle\langle 1|+H_{22}|2\rangle\langle 2|+H_{12}|1\rangle\langle 2|+h.c.$ The above condition can be satisfied by introducing proper laser frequencies in our Raman process. Specifically, in our calculations, with $a_0=4$, one group of possible experimental parameters are $B=10\textrm{G}$, $\lambda_a\approx791.56\textrm{nm}$, $\lambda_b\approx791.56\textrm{nm}$ and $\lambda_c\approx805.96\textrm{nm}$. The two-photon detuning between $\ket{1}$ and $\ket{2}$ transition is around $400\textrm{kHz}$. All the time-dependent and spatially dependent AC Stark shifts in the diagonal elements are negligibly small due to the opposite AC stark shift caused by laser $c$. The corresponding Floquet driving frequency is $\omega=2\pi\times72\textrm{kHz}$ and $\Omega_0=2\pi\times288\textrm{kHz}$. The Raman laser intensities are $I_a\approx24.2\textrm{W}/\textrm{cm}^2$, $I_b\approx605\textrm{W}/\textrm{cm}^2$ and $I_c\approx9.8\textrm{W}/\textrm{cm}^2$. Ignoring the constant term proportional to the identity matrix, we get the desired Hamiltonian, $H_d$, in the main text from the effective two-level Hamiltonian, $H_e$.

The next issue we care about is the scattering limited lifetime of the Bose gas. Using the well-developed theory in optical scattering in cold atom ensembles\cite{grimm2000optical}, the scattering rate can be written as
\begin{equation}\label{scattering rate}
\gamma_{sc}\approx\frac{\pi c^2\Gamma^2}{2\hbar\omega_0^3}(\frac{2}{\Delta_{D_2}}+\frac{1}{\Delta_{D_1}})(I_{a}+I_{b}+I_{c}),
\end{equation}
where $c$ is the speed of light, $\Gamma\approx2\pi\times 6\textrm{MHz}$ is the natural linewidth, and $\Delta_{D_1}$ and $\Delta_{D_2}$ are the one-photon detunings with respect to the $D_1$ and $D_2$ lines, respectively. By plugging in our parameters and using a Gaussian beam waist of $200\mu m$, the life time  $\tau_{sc}=1/\gamma_{sc}\approx133\textrm{ms}$, which is sufficient for experimental applications.

\end{document}